\documentclass[a4paper,12pt]{article}
\usepackage[utf8]{inputenc}
\usepackage{lmodern}
\usepackage[T1]{fontenc}
\usepackage{amsmath,amsthm,amsfonts,amssymb,bm}
\usepackage{dsfont}			% indicator function
\usepackage{enumitem}
\usepackage{graphicx}
\usepackage{booktabs}
\usepackage{xcolor}
\usepackage{authblk}
\usepackage[authoryear]{natbib}
\usepackage[right=2.5cm,left=2.5cm,top=2cm,bottom=3cm]{geometry}
\usepackage[font={small,it}]{caption}	% the options makes the font italic in floats
\usepackage[colorlinks,linkcolor=blue,citecolor=blue,urlcolor=blue]{hyperref}
\usepackage{subfig}
\usepackage{mathtools}

\usepackage{multirow}
\usepackage{makecell}
\usepackage{afterpage}
\usepackage{setspace}
\newcommand{\x}{\mathbf{x}}

\newcommand{\MU}{\bm{\mu}}
\newcommand{\SIGMA}{\bm{\Sigma}}
\newcommand{\OMEGA}{\bm{\Omega}}
\newcommand{\PSI}{\bm{\Psi}}
\newcommand{\THETA}{\bm{\Theta}}

\DeclareMathOperator*{\argmax}{arg\,max}
\newcommand{\norm}[1]{\left\lVert#1\right\rVert}

\title{ \textsc{Group-wise shrinkage estimation in penalized model-based clustering} }

\author[1]{Alessandro Casa\footnote{These authors contributed equally to this work. \\ \hspace*{0.33cm} $^1$ \hspace*{-0.2cm} Corresponding author: Faculty of Economics and Management, Free University of Bozen-Bolzano, Piazza Università 1, 39100 Bolzano, Italy. Email: \texttt{alessandro.casa@unibz.it}}}
\author[2]{Andrea Cappozzo$^*$}
\author[3]{Michael Fop}
\affil[1]{Faculty of Economics and Management, Free University of Bozen-Bolzano}
\affil[2]{MOX - Laboratory for Modeling and Scientific Computing, Politecnico di Milano}
\affil[3]{School of Mathematics \& Statistics, University College Dublin}

\date{}                     %% if you don't need date to appear

\begin{document}

\maketitle

\begin{abstract}
Finite Gaussian mixture models provide a powerful and widely employed probabilistic approach for clustering multivariate continuous data.
However, the practical usefulness of these models is jeopardized in high-dimensional spaces, where they tend to be over-parameterized. As a consequence, different solutions have been proposed, often relying on matrix decompositions or variable selection strategies. Recently, a methodological link between Gaussian graphical models and finite mixtures has been established, paving the way for penalized model-based clustering in the presence of large precision matrices. Notwithstanding, current methodologies implicitly assume similar levels of sparsity across the classes, not accounting for different degrees of association between the variables across groups.
We overcome this limitation by deriving group-wise penalty factors, which automatically enforce under or over-connectivity in the estimated graphs. The approach is entirely data-driven and does not require additional hyper-parameter specification. Analyses on synthetic and real data showcase the validity of our proposal.
%not accounting for the fact that groups may possess different degrees of association among the variables. 
%, and they have been employed in a plethora of applications.

%Gaussian Graphical Models are widely employed for modelling dependence among variables. Likewise, finite Gaussian mixtures are often the standard way to go for model-based clustering of continuous features. With the increasing availability of high-dimensional datasets, a methodological  link between these two approaches has been established in order to provide a framework for performing penalized model-based clustering in the presence of large precision matrices. Notwithstanding, current methodologies do not account for the fact that groups may possess different degrees of association among the variables, thus implicitly assuming similar levels of sparsity across the classes. We overcome this limitation by deriving group-wise penalty factors, automatically enforcing under or over-connectivity in the estimated graphs. The approach is entirely data-driven and does not require any additional hyper-parameter specification. Simulated data experiments showcase the validity of our proposal.
\end{abstract}


%-------------------------------------------------------------------------
\section{Introduction}\label{sec:introduction}
In their recent work, \citet{gelman:2020} include regularized estimation procedures among the most important contributions to the statistical literature of the last fifty years. Technological advancements and the booming of data complexity, both from a dimensional and structural perspective, have fostered the development of complex models, often involving an increasingly large number of parameters. Different regularization strategies have been proposed to obtain good estimates and predictions in these otherwise troublesome settings. In this framework, a considerable amount of effort has been put into the estimation of sparse structures \citep[see][for a review]{hastie:2015}. The rationale underlying the sparsity concept assumes that only a small subset of parameters of a given statistical model is truly relevant. As a consequence, sparse procedures usually include penalization terms in the objective function to be optimized, forcing the estimates of some parameters to be equal to zero. These machineries generally lead to an improvement in terms of interpretability and stability of the results, as well as to advantages from a computational perspective, while reducing the risk of overfitting. 

Sparse modelling has been successfully applied in regression and in supervised classification contexts. Furthermore, these strategies have been recently employed also in the model-based clustering framework, where Gaussian mixture models are usually considered to group multivariate continuous data. As a matter of fact, these models tend to be over-parameterized in high-dimensional scenarios \citep[see][for a discussion]{bouveyron2014model}, where the detection of meaningful partitions becomes more troublesome. For this reason, penalized likelihood methods have been considered, inducing sparsity in the resulting parameter estimates, and possibly performing variable selection \citep[see e.g.][]{Pan2007,xie2008penalized,zhou2009penalized}. In particular \citet{zhou2009penalized} propose a penalized approach which drastically reduces the number of parameters to be estimated in the component inverse covariance, or precision, matrices. This method exploits the connection between Gaussian mixture models and Gaussian graphical models \citep[GGM,][]{whittaker:1990}, which provides a convenient way to graphically represent the conditional dependencies encoded in the precision matrices. The estimation of such matrices is difficult when the number of variables is comparable to or greater than the sample size. For this reason, a fruitful line of research has focused on sparsity inducing procedures, which allows to obtain estimates in high-dimensional scenarios: readers may refer to \citet{Pourahmadi2013} for a detailed treatment of the topic.

The approach by \citet{zhou2009penalized} induces the intensity of the penalization imposed to be common for all the component precision matrices, thus implicitly assuming that the conditional dependence structure among the variables is similar across classes. This assumption can be harmful and too restrictive in those settings where the association patterns are cluster-dependent. For instance, the method can be inappropriate to classify subjects affected by autism spectrum disorder, which might present under or over-connected fMRI networks with respect to control individuals \citep[see][for a review on the topic]{hull:2017}. Another relevant example can be found in the field of digits recognition, for which dependence structures between pixels may vary greatly across digits: a comprehensive analysis is reported in Section \ref{sec:digit_recogn}. 

In order to circumvent this drawback, a possible solution consists in considering class-specific penalization intensities. While reasonable, this approach implies a rapidly increasing computational burden and it substantially becomes impractical even with a moderate number of classes. Other viable strategies would resort to procedures that deal with the estimation of GGMs in the multi-class framework \citep[see e.g.][and references therein]{danaher:2014}. Nonetheless, most of these proposals adopt a borrowing-strength strategy, encouraging the estimated precision matrices to be similar across classes. This behaviour may be inappropriate in a clustering context, since it might hinder groups discrimination and jeopardize the output of the analysis. 

In this work, taking our step from \citet{friedman:2008} where single class inverse covariance estimation is considered, we introduce a generalization of the method by \citet{zhou2009penalized}, which may be consequently seen as a particular case of our proposal. Even if considering a single penalization parameter, thus avoiding the troublesome selection of more shrinkage terms, our approach turns out to be more flexible and adaptive since it penalizes a class-specific transformation of the precision matrices rather than the matrices themselves. In such a way, we are able to encompass under or over-connectivity situations as well as scenarios where the GGMs share similar structures among the groups.

The rest of the paper is structured as follows. Section \ref{sec:preliminaries} briefly recalls the model-based clustering framework, with a specific focus on the strategies proposed to deal with over-parameterized mixture models. In Section \ref{sec:proposal} we motivate and present our proposal, both in terms of model specification and estimation. In Sections \ref{sec:application} and \ref{sec:true_data} the performances and the applicability of the proposed approach are tested on synthetic and real data, respectively. Lastly, the paper ends with a brief discussion in Section \ref{sec:discussion}.

\section{Preliminaries and related work}\label{sec:preliminaries}

Model-based clustering \citep{Fraley2002,bouveyron:2019} represents a well established and probabilistic-based approach to account for possible heterogeneity in a population. % and to cluster multivariate data.
In this framework, the data generating mechanism is assumed to be adequately described by means of a finite mixture of probability distributions, with a one-to-one correspondence between the mixture components and the unknown groups. More specifically, let $\mathbf{X} = \{\x_1, \dots, \x_n \}$ be the set of observed data with $\x_i \in \mathbb{R}^p$, for $i = 1, \dots, n$, and $n$ denoting the sample size. The density of a generic data point $\x_i$ is given by 
\begin{eqnarray}\label{eq:mixture}
f(\x_i ; \bm{\Psi}) = \sum_{k=1}^K \pi_k f_k(\x_i; \bm{\Theta}_k)
\end{eqnarray}  
where $K$ is the number of mixture components, $\pi_k$'s are the mixing proportions with $\pi_k > 0$ and $\sum_k \pi_k =1$, and $\PSI = \{\pi_1, \dots, \pi_{K-1}, \THETA_1, \dots, \THETA_k \}$ is the vector of model parameters. 

In (\ref{eq:mixture}), $f_k(\cdot; \THETA_k)$ represents the generic $k$-th component density; even if other flexible choices have been proposed \citep[see, e.g.][]{mclachlan1998robust,lin2009maximum,lin2010robust,vrbik2014parsimonious}, when dealing with continuous data, Gaussian densities are commonly employed. Therefore, we assume that $f_k(\cdot; \THETA_k) = \phi(\cdot; \MU_k, \SIGMA_k)$, where $\phi(\cdot; \MU_k, \SIGMA_k)$ denotes the density of a multivariate Gaussian distribution with mean vector $\MU_k = (\mu_{1k},\dots,\mu_{pk} )$, covariance matrix $\SIGMA_k$, and with $\THETA_k = \{\MU_k, \SIGMA_k\}$, for $k=1,\dots,K$.

Operationally, maximum likelihood estimation of $\bm{\Psi}$ is carried out by means of the EM-algorithm \citep{dempster1977maximum}. This is achieved by resorting to the missing data representation of model (\ref{eq:mixture}), with $\mathbf{y}_i = (\x_i, \mathbf{z}_i)$ denoting the complete data, with $\mathbf{z}_i = (z_{i1}, \dots, z_{iK})$ latent group indicators where $z_{ik} = 1$ if the $i$-th observation belongs to the $k$-th cluster and $z_{ik} = 0$ otherwise. 
Considering a one--to--one correspondence between clusters and mixture components, as it is common in the general model-based clustering framework, the partition is obtained assigning the $i$-th observation to cluster $k^*$ if 
\begin{eqnarray*}
k^* = \argmax_{k=1,\dots,K} \frac{\pi_k \phi(\x_i; \MU_k, \SIGMA_k)}{\sum_{v=1}^K \pi_{v} \phi(\x_i; \MU_{v}, \SIGMA_{v})} \; ,
\end{eqnarray*}
according to the so called maximum a posteriori (MAP) classification rule \citep[see Ch. 2.3 in][for details]{bouveyron:2019}. 

One of the major limitations of Gaussian mixture models is given by their tendency to be over-parameterized in high-dimensional scenarios. In fact, the cardinality of $\bm{\Psi}$ is of order $\mathcal{O}(Kp^2)$, thus scaling quadratically with the number of the observed variables and often being larger than the sample size. In order to mitigate this issue, several different approaches have been studied, and readers may refer to \cite{bouveyron2014model} and \citet{fop2018variable} for exhaustive surveys on the topic. Hereafter,  we outline some of the proposals introduced to deal with over-parameterized mixture models. Roughly speaking, we might identify three different types of approaches, namely constrained modelling, sparse estimation, and variable selection. 

The first strategy relies on constrained parameterizations of the component covariance matrices. The proposals by \citet{banfield1993model} and \citet{celeux1995gaussian} aim to reduce the number of free parameters by considering an eigen decomposition of $\SIGMA_k$, which allows to control the shape, the orientation, and the volume of the clusters. Other works falling within this framework are, to mention a few, the ones by \citet{mclachlan2003modelling,mcnicholas2008parsimonious,bouveyron2007high} and \citet{biernacki2014stable}. Most of these methodologies do not directly account for the associations between the observed variables, resorting to matrix decompositions and focusing on the geometric characteristics of the component densities. As a consequence, parsimony is induced in a rigid way and the interpretation in some cases is not straightforward.

The second class of approaches employs flexible sparsity-inducing procedures, to overcome the limitations of constrained modelling. As an example, we mention the methodology recently proposed by \citet{fop:2019},
%Secondly, to overcome limitations of constrained modeling, more flexible sparsity inducing procedures may be employed. On this wise, we mention the methodology recently proposed by \citet{fop:2019}, 
%To overcome such limitations, a sparsity inducing estimation procedure has been recently proposed by \citet{fop:2019}
where a mixture of Gaussian covariance graph models is devised, coupled with a penalized likelihood estimation strategy. This approach eases the interpretation of the results in terms of marginal independence among the variables and allows for cluster-wise different association structures, by obtaining sparse estimates of the covariance matrices.

Lastly, variable selection has been explored in this context, following two distinct paths. On one hand, the problem has been recast in terms of model selection, with models defined considering different sets of variables being compared by means of information criteria \citep{raftery2006variable,maugis2009variable,maugis2009variable_2}. On the other hand, the second class of approaches lies in between the variable selection and the sparse estimation methodologies. In fact, in \citet{Pan2007,xie2008penalized,zhou2009penalized} a penalty term is considered in the Gaussian mixture model log-likelihood to induce sparsity in the resulting estimates and thus possibly identifying a subset of irrelevant variables. 

\iffalse
In the following, we focus specifically on the work by \citet{zhou2009penalized}, where penalties are placed on the mean and inverse covariance parameters of the Gaussian mixture components. The penalty on the cluster-specific means is employed to perform variable selection, while the penalty on the component precision matrices is considered for regularization and to obtain sparse estimates of the association matrices in high-dimensional settings. Parameter estimation and the subsequent clustering step are carried out by maximizing the following penalized log-likelihood: 
% \begin{eqnarray}\label{eq:eq2}
% \ell^Z_P(\bm{\Psi}) = \sum_{i=1}^n \log \sum_{k=1}^K \pi_k \phi(\x_i; \MU_k, \OMEGA_k) - \lambda_1\sum_{k=1}^K \sum_{j=1}^p |\mu_{jk}| - \lambda_2 \sum_{k=1}^K\norm{\OMEGA_k}_1,
% \end{eqnarray}
\begin{eqnarray}\label{eq:eq2}
\tilde{\ell}_P(\bm{\Psi}) = \sum_{i=1}^n \log \sum_{k=1}^K \pi_k \phi(\x_i; \MU_k, \OMEGA_k) - \lambda_1\sum_{k=1}^K \sum_{j=1}^p |\mu_{jk}| - \lambda_2 \sum_{k=1}^K\norm{\OMEGA_k}_1,
\end{eqnarray}
\fi

In the following, we focus specifically on the work by \citet{zhou2009penalized}, where the penalty is placed on the inverse covariance parameters of the Gaussian mixture components. Such a penalty is considered for regularization and for obtaining sparse estimates of the association matrices in high-dimensional settings. Parameters estimation, and the subsequent clustering step, are carried out by maximizing the following penalized log-likelihood
\begin{eqnarray}\label{eq:eq2}
\tilde{\ell}_P(\bm{\Psi}) = \sum_{i=1}^n \log \sum_{k=1}^K \pi_k \phi(\x_i; \MU_k, \OMEGA_k) - \lambda \sum_{k=1}^K\norm{\OMEGA_k}_1.
\end{eqnarray}
The first term is the log-likelihood of a Gaussian mixture model, parametrized in terms of the component precision matrices $\OMEGA_k = \SIGMA_k^{-1}$, for $k=1,\dots,K$. The second term corresponds to the graphical lasso penalty \citep[see, e.g.][]{banerjee:2008,friedman:2008,scheinberg:2010,witten:2011} applied to the component-specific precision matrices, with the $L_1$ norm taken elementwise, i.e. $\norm{A}_1 = \sum_{ij} |A_{ij}|$; in the following, we do not apply the penalty on the diagonal elements of the precision matrices, even if it is possible in principle. \\Note that \citet{zhou2009penalized} consider an additional penalty term $\lambda_2\sum_{k=1}^K \sum_{j=1}^p |\mu_{jk}|$ in \eqref{eq:eq2}: this corresponds to the lasso penalty function \citep{tibshirani:1996} applied element-wise to the mean component vectors employed to perform variable selection. Since our primary focus is in uncovering the conditional dependence structure enclosed in $\OMEGA_k$, we are not concerned in providing penalized estimators for the component mean vectors, we therefore do not include this term in \eqref{eq:eq2}.
%Therefore, with the proposed strategy, the only parameters requiring a careful tuning are $\lambda$ and the number of components $K$: 
% in Section \ref{sec:further_aspect} a standard technique to do so according to an objective criterion is outlined. } 

The graphical lasso penalty allows to induce sparsity in the precision matrices, which eases the interpretation of the model. In fact, $\OMEGA_k$ embeds the conditional dependencies among the variables for the $k$-th component, whereby in the Gaussian case zero entries between pair of variables imply that they are conditionally independent given all the others. Moreover, a convenient way to visually represent the dependence structure among the features is given by the graph of the associated Gaussian graphical model. Here, as already mentioned in the introduction, a correspondence between a sparse precision matrix and a graph is defined, with nodes representing the variables while the edges connect only those features being conditionally dependent. A recent and interesting extension is represented by the \emph{colored graphical models}, where symmetry restrictions are added to the precision matrix thus offering a more parsimonious representation and possibly highlighting commonalties among the variables; readers may refer to \cite{hojsgaard2008graphical, gao2015estimation, li2021penalized} and references therein for a more detailed discussion.

Sparse precision matrix estimation via the graphical lasso algorithm is routinely employed assuming that observations arise from the same population, adequately described by a single multivariate Gaussian distribution, which is indexed by a single GGM. However, this assumption does not hold in the cluster analysis framework where, as in \eqref{eq:mixture}, the observed data are assumed to arise from $K$ different sub-populations, which might be characterized by different association patterns. %Such scenarios require the estimation of $K$ class-specific GGMs, defined in terms of a collection of precision matrices $\Omega_1, \dots, \OMEGA_k$. 
Some modifications of the standard graphical lasso have been proposed, in order to make it applicable also in a multi-class setting \citep[see, e.g.][]{guo:2011,mohan:2014,danaher:2014,lyu:2018}. Nonetheless, these approaches usually consider the matrices $\OMEGA_k$'s to have possible commonalities and shared sparsity patterns; as a consequence, they modify the graphical lasso penalty term in order to induce the estimated GGMs to be similar to each other, while allowing for structural differences. These strategies have been usually considered with an exploratory aim in mind, in order to obtain parsimonious and interpretable characterization of the relationships among the variables within and between the classes. Undoubtedly, they might be fruitfully embedded also in a probabilistic unsupervised classification context. However, to some extent, by borrowing strength and encouraging similarity among groups, these methods may be inappropriate, if not harmful, as they might hinder the classification task itself. This particularly holds in the case of clustering, where the classes are not readily available and need to be inferred from the data. For this reason, in the next section we focus on how to obtain cluster-specific sparse precision matrices to account for cluster-wise distinct degrees of sparsity.

\section{Proposal} \label{sec:proposal}
All the multi-class GGM estimation strategies reviewed in Section \ref{sec:preliminaries} assume that different classes are characterized by a similar structure in the precision matrices, either explicitly or implicitly. This assumption is explicit for those approaches where similarities among the precision matrices are encouraged by the considered penalty term. % Nonetheless, as already briefly mentioned, embedding these strategies into a classification framework can hamper the final aim of the analysis since it could force some class-defining characteristics to be equal, or at least more similar.
Similarly, in the approach proposed by \citet{zhou2009penalized} the assumption is implicitly entailed by the use of a single penalization parameter $\lambda$. The adopted penalization scheme, even if somehow weighted by the clusters sample sizes, as per Equation \eqref{eq:max_problem_omega}, can be profitably considered only in those situations where the number of non-zero entries in the precision matrices is similar across classes. Therefore, these approaches do not contemplate under or over-connectivity scenarios, where the groups are characterized by significantly different amounts of sparsity. This constitutes a serious limitation in those applications where different degrees of connectivity could ultimately characterize the resulting data partition. That is, whilst approaches such as the one by Zhou et al. (2009) well encompass scenarios in which connected nodes are group-wise different, we aim at defining a data-driven strategy specifically designed for addressing also those situations in which groups differ in the amount of sparsity (i.e., in the number of non-zero entries in the precision matrices).
\subsection{Motivating example}
%\begin{figure}[!tbp]
%  \centering
%  \begin{minipage}[b]{0.49\textwidth}
%  	\centering
%    \includegraphics[width = 7cm, height = 7.3cm]{figures/sparsePrec_1st.pdf}
%  \end{minipage}
%  \hspace{-2cm}
%  \begin{minipage}[b]{0.49\textwidth}
%  	\centering
%    \includegraphics[width = 7cm, height = 7.3cm]{figures/sparsePrec_2nd.pdf}
%  \end{minipage} \par
%  \vspace*{-1cm}
%  \includegraphics[width=10cm, height = 5.5cm]{figures/tradeoffplot.pdf}
%  \caption{On the top: true data generating component-specific precision matrices. Black squares denote the presence of an edge between the two variables in the corresponding GGM. On the bottom: classification accuracy, estimated as the percentage of null entries in the true precision matrices being estimated as zero, varying the strength of the penalization (dashed blue line: first component, red dotted line: second component)}
%  \label{fig:fig1}
%\end{figure}
\begin{figure}
\centering
    \includegraphics[scale=.45]{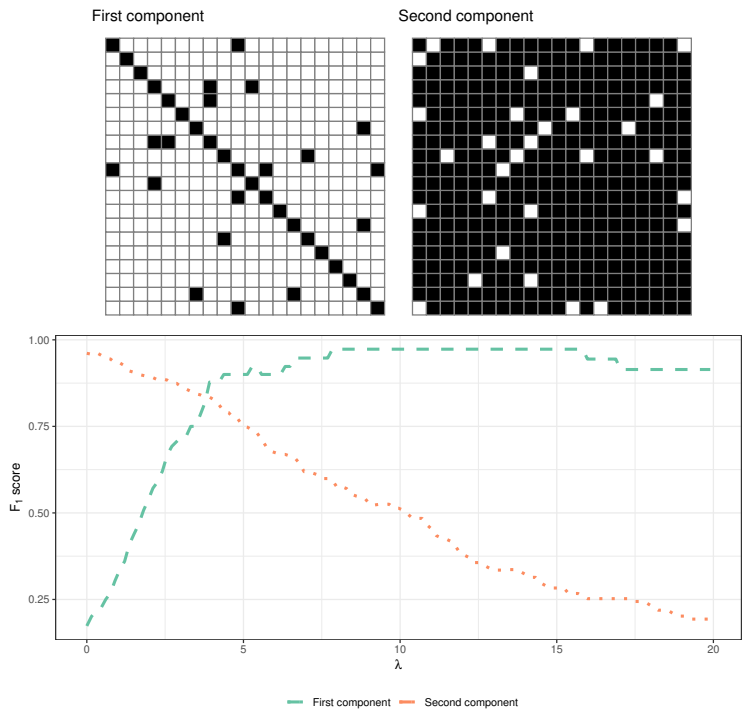}	
    \caption{On the top: true data generating component-specific precision matrices. Black squares denote the presence of an edge between the two variables in the corresponding GGM. On the bottom: $F_1$ score as a function of the parameter $\lambda$.}
  \label{fig:motivating_ex}
\end{figure}

To better explain this issue and to justify our solution, we provide a motivating example where we simulate $n=200$ \emph{p}-dimensional observations with $p=20$ from a mixture with $K = 2$ Gaussian components and mixing proportions $\pi = (0.5,0.5)$. 
%The component means are chosen to be equal, so that the clustering is only driven by the group-specific covariances. 
The considered component precision matrices are associated with the two sparse at random structures depicted in the top panel of Figure \ref{fig:motivating_ex}. %\comment{MF - Can we make this figure with the two panels side by side? It  would save a lot of space}.
The first component is characterized by an almost diagonal precision matrix while the second presents a dense structure, thus mimicking a scenario where the degree of sparsity is drastically different among the two classes. We estimate these matrices employing the penalization scheme in \eqref{eq:eq2}, with $\lambda$ ranging over a suitable interval. The ability in recovering the association patterns inherent to the two clusters is evaluated by means of the $F_1$ score:
\begin{eqnarray} \label{eq:f1_measure}
F_1=\frac{\texttt{tp}}{\texttt{tp}+0.5(\texttt{fp}+\texttt{fn})},
\end{eqnarray}
where $\texttt{tp}$ denotes the correctly identified edges (i.e., the number of non-zero entries in the precision matrix correctly estimated as such), while $\texttt{fp}$ and $\texttt{fn}$ represent respectively the number of incorrectly identified edges and the number of missed edges (i.e. the number of non-zero entries wrongly shrunk to $0$). Line plots displaying the $F_1$ patterns for the two components are reported in the bottom panel of Figure \ref{fig:motivating_ex}.
%the classification accuracy of the true zero entries in the matrices, are reported in the bottom panel of Figure \ref{fig:fig1}.
A trade-off is clearly visible, indicating how a common penalty term prevents the possibility to obtain a proper estimation of both the precision matrices. In fact, if for the second component a mild penalization might be adequate to estimate the dense dependence structure, for the first component the high degree of sparsity is recovered only when a stronger penalty is considered.   

%A trade-off is clearly visible  in the class-specific metric, preventing a proper estimation of both the precision matrices by means of a common penalty term. In fact, a mild penalization might be adequate to properly estimate the dense relationships among variables in the second component, while the sparse structure of the first component requires a stronger penalty. 
\subsection{Model specification}\label{sec:modelspec}
The illustrative example clearly shows how the penalization strategy proposed by \citet{zhou2009penalized} does not represent a suitable solution when dealing with unbalanced class-specific degrees of sparsity among the variables. As briefly mentioned in the introduction, a possible alternative would consist in uncoupling the precision matrices estimation by considering component specific penalization coefficients. That is, $\lambda$ in \eqref{eq:eq2} would be substituted by $K$ different shrinkage terms. 
%penalized log-likelihood as 
%\begin{eqnarray}\label{eq:eq4}
%\ell_P(\Theta) = \sum_{i=1}^n \log \sum_{k=1}^K \pi_k \phi(\x_i; \MU_k, \OMEGA_k) - \sum_{k=1}^K \lambda_k \norm{\OMEGA_k}_1.
%\end{eqnarray}
%The increased flexibility of this approach is then obtained by considering component specific penalization coefficients. 
While in principle reasonable, the  increased flexibility induced by introducing $K$ different penalties may be problematic, since in the graphical lasso framework tuning these hyper-parameters is a difficult task. Time consuming grid searches are generally considered, with the optimal penalty factor selected either according to some information criteria %computed on the corresponding model
or resorting to computationally intensive cross-validation strategies. The simultaneous presence of $K$ penalty terms would make this approach much more complex, also from a computational perspective. % In addition, the non-identifiability issue in mixture models (also known as label switching problem) would exacerbate the problem even further, obliging the user to fix. 

%\textcolor{blue}{AC: non si rischia di avere un problema anche relativo alla non identifiability dei mixture models?}

In this work %we take a related but different path as
we propose instead to carry out parameter estimation by maximizing a penalized log-likelihood function defined as follows:
\begin{eqnarray}\label{eq:obj_function}
\ell_P(\PSI) = \sum_{i=1}^n \log \sum_{k=1}^K \pi_k \phi(\x_i; \MU_k, \OMEGA_k) - \lambda\sum_{k=1}^K \norm{\mathbf{P}_k * \OMEGA_k}_1,
\end{eqnarray}
where $*$ denotes the element-wise product between two matrices and $\mathbf{P}_k$'s are weighting matrices employed to scale the effect of the penalty.
% \comment{MF -- Specify that $\mathbf{P}_k$ does not need to be positive definite, and that elements of $\mathbf{P}_k$ needs to be non-negative.}
% \comment{ACAS -- At a quick reading a first comment arising would be ``You do not want to have $K$ regularization parameters but you ended up having $K$ matrices to specify'' and it looks like we are complicated the thing. It is clear from the following sections that it's easier to specify these matrices but probably it's worth adding a sentence here (?) -- MF - agree, we need to phrase carefully this part, otherwise it seems that we move the problem from the $\lamdba_k$ to the $\mathbf{P}_k$}
The focus is then shifted towards the specification of $\mathbf{P}_1,\dots,\mathbf{P}_K$, which, when properly encoding information about class-specific sparsity patterns, introduces a degree of flexibility that accounts for under or over-connectivity scenarios. In \eqref{eq:obj_function} a single penalization parameter for the precision matrices is considered. As a consequence %, with the proposed strategy,
other than the model selection problem concerning the selection of the number of components $K$, we only need to carefully tune a single penalization hyperparameter $\lambda$. In Section \ref{sec:further_aspect} we outline a standard technique to choose the number of components and the penalty parameter $\lambda$ according to a model selection criterion. 
%\textcolor{red}{Lastly note that, since our primary focus is on uncovering the conditional dependence structure enclosed in $\OMEGA_k$, we are not concerned in providing penalized estimators for the mean component vectors, therefore we do not include a penalty on $\MU_k$.} 
%Therefore, with the proposed strategy, the only parameters requiring a careful tuning are $\lambda$ and the number of components $K$: 
% in Section \ref{sec:further_aspect} a standard technique to do so according to an objective criterion is outlined. 

Hereafter, we describe a data-driven procedure for inferring $\mathbf{P}_k$'s by means of carefully initialized sample precision matrices $\hat{\OMEGA}^{(0)}_1,\dots,\hat{\OMEGA}^{(0)}_K$. Our proposals rely on the definition of a function $f:\mathbb{S}^p_+\rightarrow \mathbb{S}^p$, where $\mathbb{S}^p_+$ and $\mathbb{S}^p$ respectively denote the space of positive semi-definite and symmetric matrices of dimension $p$, such that $\mathbf{P}_k = f(\hat{\OMEGA}^{(0)}_k)$.

Recommendations on how to compute $\hat{\OMEGA}^{(0)}_k$, and how to subsequently define $\mathbf{P}_k$,  $k=1,\ldots,K$ will be the object of the next subsections. 
% by means of carefully initialized sample precision matrices $\hat{\Omega}^{(0)}_1,\dots,\hat{\Omega}^{(0)}_K$, as described in the next subsections

%making the hyper-parameter specification%computational part more manageable. 
%FIXME where to put this? 

\subsection{Initializing the sample precision matrices $\hat{\OMEGA}^{(0)}_k$'s} \label{sec:init_omega_0}
The definition of a proper strategy to initialize the matrices $\hat{\OMEGA}^{(0)}_k$'s, for $k=1,\dots,K$, strongly depends on the framework and on the specific task of the analysis. In fact, in supervised and semi-supervised scenarios, where the class labels are known for at least a subset of observations, the initialization step turns out to be straightforward. More formally, let us denote with $n$ the number of observations in the training set, with $m$ the number of observations with known labels and $m=\sum_{k=1}^K m_k$, with $m_1, \dots, m_K$ denoting the class specific sample sizes with $m_k > 0$, for $k=1,\dots,K$. In the supervised setting, where $m=n$, and in the semi-supervised one, where $0 < m < n$, we simply define $\hat{\OMEGA}^{(0)}_k$ to be the $k$-th class sample precision matrix estimated on the pertaining $m_k$ observations. Note that, if $p > m_k$, the initialized sample precision matrix might be obtained by means of $K$ distinct graphical lasso algorithms. 

On the other hand, in a clustering framework where $m=0$, the specification of $\hat{\OMEGA}^{(0)}_k$ is more tricky. The absence of clear indications about the group memberships makes the approach introduced for the supervised and semi-supervised context impractical. Nonetheless, it is possible to find suitable workarounds in order to employ our proposal even in an unsupervised scenario.
%to extend our proposal to the clustering framework.
From a practical point of view, we consider a general multi-step procedure as follows:
\begin{enumerate}
	\item Run any clustering algorithm on the observed data $\mathbf{X} = \{\x_1,\dots,\x_n \}$, to obtain an initial partition of the observations into $K$ groups;
	%$\mathbf{Z}^{(0)}=\{z_1,\dots,z_n\}$ with $z_i = k$ if the $i$-th observation belongs to $k$-th cluster having cardinality equal to $n_k$ \\
	\item Given the obtained initial partition, estimate the cluster specific precision matrices $\hat{\OMEGA}^{(0)}_1, \dots,\hat{\OMEGA}^{(0)}_K$ using only those observations assigned to that specific group; again $\hat{\OMEGA}^{(0)}_k$ might be obtained as the sample estimate when $p<n_k$ or as the graphical lasso solution if $p \ge n_k$.
%	\item Specify the matrices $\mathbf{P}_k$'s as an adequately chosen function of the cluster specific precision matrices, i.e. $\mathbf{P}_k = f(\hat{\Omega}^{(0)}_k)$, for $k = 1,\dots, K$.  
\end{enumerate}
When resorting to this procedure, the choice of which clustering strategy to adopt for obtaining the initial partition is subjective and needs to be carefully taken. In fact, inadequate choices might provide incoherent indications of the true clustering structure and hinder the possibility %to subsequently recover it and
to obtain an accurate reconstruction of the dependencies among the variables when maximizing \eqref{eq:obj_function}. In principle, different clustering strategies may be adopted, and providing specific suggestions about the more adequate ones is beyond the scope of this work. Nonetheless, model-based techniques \citep[see e.g.][]{Fraley2002} might constitute a clever choice, being them coherent with the considered framework. Also ensemble strategies can be adequate, as they aim to combine the strengths of different algorithms and lessen the impact of some otherwise cumbersome choices \citep[see][for some proposals from a model-based standpoint]{russell2015bayesian,wei2015mixture,Casa2020}. In addition, powerful initialization strategies for partitioning the data into $K$ groups  can as well appropriately serve the purpose \citep{Scrucca2015}. Lastly, we remark that the use of standard methods to obtain an appropriate initial clustering partition can cause difficulties in high-dimensional settings where $p>n$. In these scenarios, subspace clustering methods specifically designed for high-dimensional data can be employed. Examples are mixtures of factor analyzers \citep{mcnicholas2008parsimonious} and model-based discriminant subspace clustering \citep{bouveyron2012simultaneous}; we point the reader to Section 5 of \cite{bouveyron2014model} for a comprehensive overview.

\subsection{Obtaining $\mathbf{P}_k$'s  via inversely weighted sample precision matrices} \label{sec:weigthed_by_w0}
% \comment{ACAS -- If we want to save some space (not necessarily needed, max length is 10pages without ref) we could have a single subsection here, incorporating 2.2.2, 2.2.3 and 2.2.4 that are quite short, thus saving the multiple titles space} \\
Once the sample precision matrices $\hat{\OMEGA}^{(0)}_k$'s have been initialized, the first viable proposal for defining $\mathbf{P}_k$'s reads as follows:
\begin{eqnarray} \label{weigthed_by_w0}
P_{k,ij} = 1/\left(|\hat{\Omega}^{(0)}_{k,ij}|\right), \quad \forall i,j=1,\ldots,p,
\end{eqnarray}
where $P_{k,ij}$,  $\hat{\Omega}^{(0)}_{k,ij}$ are respectively the $(i,j)$-th elements of the matrices $\mathbf{P}_k$ and $\hat{\OMEGA}^{(0)}_{k}$. Notice that it is sufficient to set $P_{k,ii}=0$, $\forall i=1,\ldots,p$, whenever the diagonal entries of $\OMEGA_k$ shall not be penalized. Intuitively, with \eqref{weigthed_by_w0} we are inflating/deflating the penalty enforced on the $(i,j)$-th element of the matrix $\OMEGA_k$ according to the magnitude of $\hat{\Omega}^{(0)}_{k,ij}$. Clearly, values of $|\hat{\Omega}^{(0)}_{k,ij}|$ close to $0$ induce a higher penalty on $\Omega_{k,ij}$. Should $\hat{\OMEGA}^{(0)}_k$ be estimated via the graphical lasso, e.g., in those situations where $p \geq n_k$, a small positive constant is added in the denominator of \eqref{weigthed_by_w0} to avoid having an undefined  $\mathbf{P}_k$.
%the matrices $\mathbf{P}_k$'s might be defined as  with zero entries on the diagonal and with $P_{k,ij}$ being the $(i,j)$-th element of the matrix $\mathbf{P}_k$. Note that $\hat{\Omega}^{(0)}_k$, for $k=1,\dots,K$, might be obtained as the sample inverse covariance matrices when $p < n$ or as the solution of the standard graphical lasso algorithm, when $ p \ge n$.
 This strategy shares connections with the proposal by \citet{bien:2011} developed in a covariance estimation context, and it might be seen as a multiclass extension of the approach proposed in \citet{fan:2009}, where the \emph{adaptive lasso} \citep{zou:2006} is generalized to the estimation of sparse precision matrices. %Other specifications of the matrices $\mathbf{P}_k$'s might be considered, also depending on possibly available prior information and on the problem at hand. An interesting choice, that showed good performances in practice, might consist in defining the off-diagonal element of the matrices as follows 
 
An hard-thresholding version of \eqref{weigthed_by_w0} may also be considered, in which entries of $\OMEGA_k$ are not shrunk if their magnitude exceeds a given constant. A sensible way to do so would be to examine  the initialized partial correlation matrix for the $k$-th class, and to fix a value $\gamma \in (0,1)$ that acts as a user-defined threshold. 
%\begin{eqnarray}\label{eq:weigthed_by_w0_thresh}
%P_{k,ij} = 
%\begin{cases} 
%1/\left(\varepsilon+ |\hat{\Omega}^{(0)}_{k,ij}|\right) & \mbox{if } |\hat{\Gamma}^{(0)}_{k,ij}| < \gamma \\ 
%\hspace*{0.4cm} 0 & \mbox{otherwise},
%\end{cases}
%\end{eqnarray}
%where $\hat{\Gamma}^{(0)}_k$ is the initialized partial correlation matrix for the $k$-th class, and $0 \leq \gamma \leq 1$ is a positive threshold to be specified.
%\comment{MF - Not implemented in the comparison -- add sentence that is not implemented because the choice is then move to selecting an appropriate threshold $\gamma$, or remove entirely?}
This approach is related to the fixed-zero problem of \citet{chaudhuri:2007}: when $\lambda$ is sufficiently large, it leads to an estimate equivalent to the one obtained from a given association graph where the zero entries correspond to partial correlations of magnitude lower than the specified threshold $\gamma$. The idea is connected to the thresholding operator for sparse covariance matrix estimation \citep{bickel:2008,Pourahmadi2013} and it has been explored in the \texttt{covglasso} R package \citep{fop:2020}. The hard-thresholded approach is not further considered in the following as it requires a sensitive choice of $\gamma$, whereas the other suggested proposals do not rely on any hyper-parameter specification.  

\subsection{Obtaining $\mathbf{P}_k$'s via distance measures in the $\mathbb{S}^p_+$ space} \label{sec:weigthed_by_distance_to_diagW0}
An alternative approach consists in setting the elements of $\mathbf{P}_k$ proportional to the distance between $\hat{\OMEGA}^{(0)}_{k}$ and ${\rm diag}\left(\hat{\OMEGA}^{(0)}_{k}\right)$, where ${\rm diag}\left(\hat{\OMEGA}^{(0)}_{k}\right)$ indicates a diagonal matrix whose diagonal elements are equal to the ones in $\hat{\OMEGA}^{(0)}_{k}$. We propose to compute $P_{k,ij}$ as follows:
\begin{eqnarray} \label{weigthed_by_distance_to_diagW0}
P_{k,ij} = \frac{1}{\mathcal{D}\left(\hat{\OMEGA}^{(0)}_k, \rm{diag}\left(\hat{\OMEGA}^{(0)}_{k}\right)\right)}, \quad \forall i,j=1,\ldots,p \quad \text{and} \quad i \neq j,
\end{eqnarray}
with $\mathcal{D}(\cdot,\cdot)$ being a suitably chosen measure of distance between positive semi-definite matrices. Since $\mathbb{S}^p_+$ is a non-Euclidean space, the type of distance needs to be carefully selected:
%several distances may be employed
the reader is referred to \citet{Dryden2009} for a thorough dissertation on the topic. Alternatively, to account for the diagonal elements in the definition of $\mathbf{P}_k$, one may employ the following quantity:   
\begin{eqnarray} \label{weigthed_by_distance_to_diagI}
P_{k,ij} = \frac{1}{\mathcal{D}\left(\hat{\OMEGA}^{(0)}_k, \mathbf{I}_p \right)}, \quad \forall i,j=1,\ldots,p \quad \text{and} \quad i \neq j,
\end{eqnarray}
where $ \mathbf{I}_p$ denotes the identity matrix of dimension $p$. The definitions of \eqref{weigthed_by_distance_to_diagW0} and \eqref{weigthed_by_distance_to_diagI} simply stems from the conjecture that the ``closer'' $\hat{\OMEGA}^{(0)}_{k}$ is to a diagonal matrix, the higher the group-wise penalty shall be, thus forcing some of the entries in $\OMEGA_{k}$ to be shrunk to $0$.

\subsection{Obtaining $\mathbf{P}_k$'s: comparison of methods} 
The strategies mentioned above share the same underlying rationale as they aim to impose stronger penalization on those entries corresponding to weaker sample conditional dependencies among variables. Moreover, being the specification class specific, they fruitfully encompass situations where one or more groups present different sparsity patterns and magnitudes.  While the solution in Section \ref{sec:weigthed_by_w0} induces an entry-wise different penalty, it heavily depends on the estimation of $\hat{\OMEGA}^{(0)}_k$. Therefore, in those situations where the reliability of the sample estimates is difficult to assess, it might be convenient to let $\mathbf{P}_k$s depend on a group specific constant, as for the strategy in Section \ref{sec:weigthed_by_distance_to_diagW0}. 

The proposed approaches generalize the one by \citet{zhou2009penalized}, as the strategies coincide when $\mathbf{P}_k$ is set to be equal to a matrix of ones for all  $k=1,\dots,K$.
%accounting for possible under or over-connectivity scenarios. In fact, such original procedure is retrieved setting $\mathbf{P}_k$ equal to the all-one matrix for all  $k=1,\dots,K$. 
 Once the definition of $\mathbf{P}_k$ has been established, coherently to \citet{zhou2009penalized}, the model is estimated employing an EM algorithm: details are given in the next section.
  
\subsection{Model estimation}
For a fixed number of components $K$ and penalty terms $\lambda$ and $\mathbf{P}_k$, model estimation deals with the maximization of \eqref{eq:obj_function} with respect to $\PSI$. Within the EM framework, a \textit{penalized complete-data log-likelihood} is naturally defined as follows:
\begin{eqnarray}\label{eq:comp_obj_function}
\ell_{C}(\PSI) = \sum_{i=1}^n\sum_{k=1}^K z_{ik} \log  \pi_k \phi(\x_i; \MU_k, \OMEGA_k) - \lambda\sum_{k=1}^K \norm{\mathbf{P}_k * \OMEGA_k}_1,
\end{eqnarray}
where as usual $z_{ik}$ is the categorical latent variable indicating the component which observation $\x_i$ belongs to. The algorithm alternates between two steps. At the $t$-th iteration, the E-step provides the expected value $\hat{z}^{(t)}_{ik}$ for the unknown labels $z_{ik}$ given $\hat{\PSI}^{(t-1)}$ , while in the M-step \eqref{eq:comp_obj_function} is maximized to determine $\hat{\PSI}^{(t)}$, conditioning on $\hat{z}^{(t)}_{ik}$.

In details, in the E-step the posterior probability of $\x_i$ belonging to component $k$ is updated  as follows:
\begin{equation}
\hat{z}^{(t)}_{ik}=\frac{\hat{\pi}_{k}^{(t-1)} \phi\left(\x_i; \hat{\MU}_k^{(t-1)}, \hat{\OMEGA}_k^{(t-1)}\right)}{\sum_{v=1}^K \hat{\pi}_{v}^{(t-1)} \phi\left(\x_i; \hat{\MU}_v^{(t-1)}, \hat{\OMEGA}_v^{(t-1)}\right)}, \quad i=1,\ldots,n.
\end{equation}
In the M-step, the updating formulas for mixing proportions and cluster means are readily given by:
\begin{equation}
\hat{\pi}_{k}^{(t)}=\frac{n_{k}^{(t)}}{n}, \quad \hat{\MU}_{k}^{(t)}=\frac{1}{n_{k}^{(t)}} \sum_{i=1}^{n} \hat{z}_{i k}^{(t)} \x_{i}
\end{equation}
where $n_{k}^{(t)}=\sum_{i=1}^n \hat{z}_{i k}^{(t)} $. Notice that, as mentioned in Section \ref{sec:preliminaries}, we are not concerned in providing penalized estimators for $\MU_k$. At any rate, should sparse mean vectors be of interest, an extra penalty 
%an objective function in the form of \eqref{eq:eq2}
can be promptly accommodated  by adding the term  $\lambda_2\sum_{k=1}^K \sum_{j=1}^p |\mu_{jk}|$ in \eqref{eq:obj_function}. In this case, estimation of sparse $\MU_k$ follows directly the steps outlined in Section 2.3.1 of  \citet{zhou2009penalized}.

When \eqref{eq:comp_obj_function} is maximized with respect to $\OMEGA_k$, the penalized complete log-likelihood simplifies as follows:
\begin{equation} \label{eq:obj_omega}
Q_{\Omega}(\OMEGA_k)=\sum_{i=1}^n \hat{z}_{i k}^{(t)} \left\lbrace\frac{1}{2} \log  \operatorname{det}(\OMEGA_k)-\frac{1}{2}  \left(\x_{i}-\hat{\MU}_{k}^{(t)}\right)^{\prime} \OMEGA_k \left(\x_{i} -\hat{\MU}_{k}^{(t)}\right) \right\rbrace - \lambda||\mathbf{P}_k * \OMEGA_k||_1.
\end{equation}
By rearranging terms in \eqref{eq:obj_omega},  the following optimization problem needs to be solved to obtain the estimate $\hat{\OMEGA}_k^{(t)}$:
\begin{equation} \label{eq:max_problem_omega}
\underset{\OMEGA_k}{\text{arg max}} \quad  \log \operatorname{det}(\OMEGA_k)-\operatorname{tr}(\mathbf{S}_k \OMEGA_k)- \frac{2\lambda}{n_{k}^{(t)}} \norm{\mathbf{P}_k * \OMEGA_k}_1,
\end{equation}
with the constraint that $\OMEGA_k$ must be positive definite, $\OMEGA_k \succ 0$, and $\mathbf{S}_k$ denoting the weighted sample covariance matrix for cluster $k$:
\begin{equation*}
\mathbf{S}_{k}=\sum_{i=1}^{n} \hat{z}_{i k}^{(t)} \frac{\left(\x_{i}-\hat{\MU}_{k}^{(t)}\right)^{\prime} \left(\x_{i}-\hat{\MU}_{k}^{(t)}\right)}{n_{k}^{(t)}}.
\end{equation*}
Following \cite{zhou2009penalized}, a coordinate descent graphical lasso algorithm by \cite{friedman:2008} is employed for solving the maximization problem in \eqref{eq:max_problem_omega}, where in our context the penalty is equal to  $2\lambda \mathbf{P}_k / n_k^{(t)}$. 

\subsection{Further aspects} \label{sec:further_aspect}
Hereafter, we discuss some practical considerations related to the algorithm devised for maximizing \eqref{eq:obj_function} and  described in the previous section.\\

%\begin{itemize}
%\item \textbf{Algorithm convergence:}
%\item \textbf{Model selection:}
%\item \textbf{Implementation:}
\noindent \textbf{Convergence:} the EM algorithm is considered to have converged once the relative difference in the objective function for two subsequent iterations is smaller than $\varepsilon$, for a given $\varepsilon>0$:
\[\frac{|\ell_{P}(\hat{\PSI}^{(t+1)})-\ell_{P}(\hat{\PSI}^{(t)})|}{|\ell_{P}(\hat{\PSI}^{(t)})|}<\varepsilon.\]
 In our analyses, $\varepsilon$ is set equal to $10^{-5}$.\\

\noindent \textbf{Model selection:} whilst the determination of $\mathbf{P}_k$s is entirely data-driven and does not require any external tuning, model selection still needs to be performed when it comes to identify the best number of components $K$ and the common penalty term $\lambda$. We rely on previous results \citep{Pan2007, Zou2007a, lian2011} which propose to select $\lambda$ and $K$ by maximizing a modified version of the Bayesian Information Criterion \citep[BIC,][]{Schwarz1978}:
\begin{eqnarray} \label{eq:modified_BIC}
B I C=2 \log L(\hat{\PSI})-d_{0}\log (n), 
\end{eqnarray}
where $\log L(\hat{\PSI})$ is the log-likelihood evaluated at $\hat{\PSI}$, obtained maximizing \eqref{eq:obj_function}, and $d_{0}$ is the number of parameters that are not shrunk to $0$ by the penalized estimation.\\
%Providing a solution on how to select the common penalty term goes beyond the scope of the present manuscript, and 

\noindent \textbf{Implementation:} routines for fitting the group-wise shrinkage method for model-based clustering have been implemented in \texttt{R} \citep{RCoreTeam2020}, and the source code is freely available at \texttt{https://github.com/AndreaCappozzo/sparsemix} in the form of an \texttt{R} package. Despite not being explicitly used in the present manuscript, the \texttt{sparsemix} software allows for penalizing the mean vectors $\MU_k$'s along the lines of \citet{zhou2009penalized}, thus providing a complete generalization of the methodology described therein.\\
%\end{itemize}
%a suitable modification accounting for the group specific penalty term. More specifically here $\lambda_{\text{gl}} = 2\lambda \mathbf{P}_k / n_k^{(t)}$ with $n_k^{(t)}$ defined as at the end of Section \ref{sec:preliminaries}.

%Lastly note that, the primary aim of this work is to throw light on a shortcoming of the available multiclass precision matrix approaches and to propose a viable solution to overcome it. Nonetheless we acknowledge that other strategies to specify $\mathbf{P}_k$'s might be fruitfully adopted. More specifically, we think that the specification should not leave aside prior information and subject-matter knowledge, when available, as well as the characteristics of the data to be analyzed. 

\begin{figure}[t]
\centering
\vspace*{-2cm}
    \includegraphics[scale=.5]{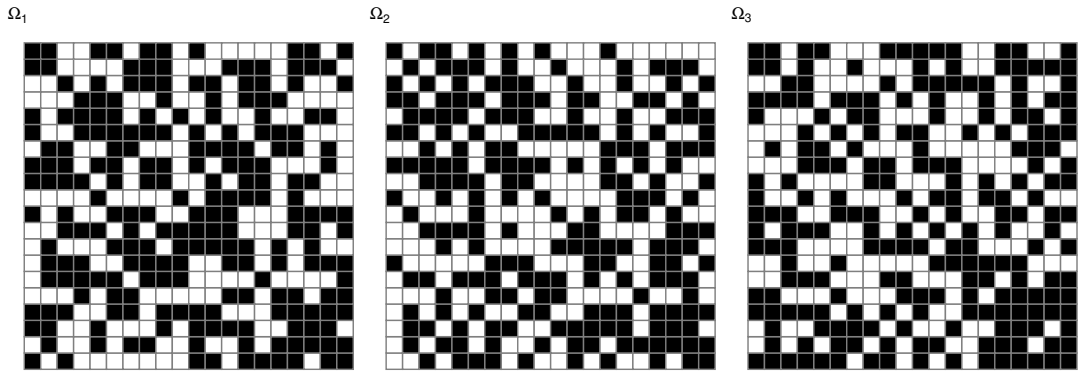}
    \vspace*{-2cm}
    \caption{Example of simulation setting \textit{Equal proportion of edges in $\OMEGA_k$}. Black squares denote the presence of an edge between the two variables.}
  \label{fig:same_p}
\end{figure}
\begin{figure}[!h]
\centering
    \vspace*{-2.5cm}
    \includegraphics[scale=.5]{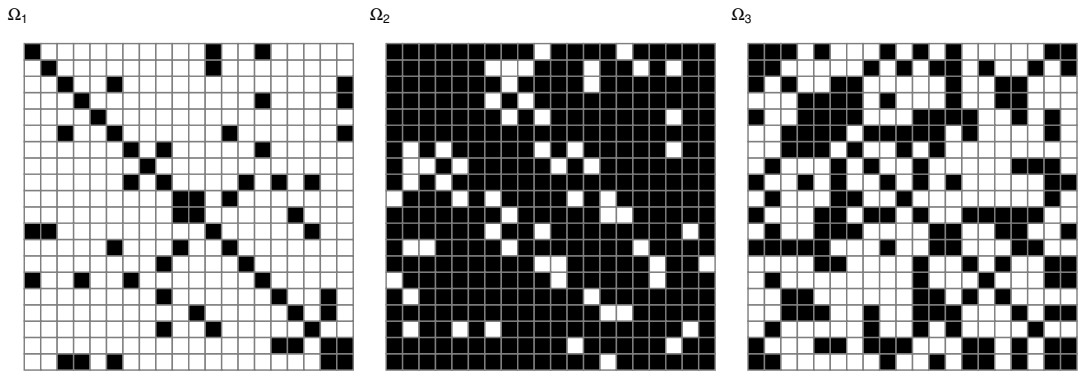}
        \vspace*{-2cm}
    \caption{Example of simulation setting \textit{Different proportion of edges in $\OMEGA_k$}. Black squares denote the presence of an edge between the two variables.}
  \label{fig:different_p}
\end{figure}
\noindent Promising results are obtained when performing penalized model-based clustering with $\mathbf{P}_k$'s defined as in \eqref{weigthed_by_distance_to_diagW0} and \eqref{weigthed_by_distance_to_diagI}, as reported in the next section.

\section{Simulated data experiment}\label{sec:application}
\subsection{Experimental setup} \label{sec:exp_setup}
We illustrate, via numerical experiments, the effectiveness of the proposed procedures in recovering the true group-wise conditional structure in a multi-class population. 
% Consider the following data generating process (DGP) for a generic observation $\x_i$:
% \begin{eqnarray*} \label{eq:sim_study_DGP}
% p(\x_i)=\sum_{k=1}^3 \pi_k \phi_p(\x_i; \MU_k, \OMEGA_k).
% \end{eqnarray*}
% According to the considered DGP $\x_i$, $i=1,\ldots,1500$, is sampled according to a $3$-component mixture model whose parameters vary according to three different scenarios:
We generate $n = 1500$ observations from a Gaussian mixture distribution with $K=3$ components, with the precision matrices $\OMEGA_k$ having various sparsity patterns, embedding different association structures. Three different scenarios are considered:
\begin{itemize}
\item \textit{Equal proportion of edges in $\OMEGA_k$}: for each replication of the simulated experiment, the precision matrices $\OMEGA_k$ are generated according to a sparse at random Erd\H{o}s-R\'{e}nyi graph structure \citep{erdHos1960evolution} characterized by the same probabilities of connection, equal to $0.5$. The number of variables is $p=20$.
\item \textit{Different proportion of edges in $\OMEGA_k$}: for each replication of the simulated experiment, the precision matrices $\OMEGA_k$ are generated according to a sparse at random Erd\H{o}s-R\'{e}nyi graph structure, with probabilities of connection equal to $0.1$, $0.8$ and $0.4$ for $\Omega_1$, $\Omega_2$ and $\Omega_3$, respectively. The number of variables is $p=20$.
\item \textit{High-dimensional and different proportion of edges in $\OMEGA_k$}: for each replication of the simulated experiment, the precision matrices $\OMEGA_k$ are generated according to a sparse at random Erd\H{o}s-R\'{e}nyi graph structure with different probabilities of connection as per the previous scenario. The number of variables is $p=100$.
\end{itemize}
\begin{figure}
\centering
%    \vspace*{-2.5cm}
    \includegraphics[scale = 0.55]{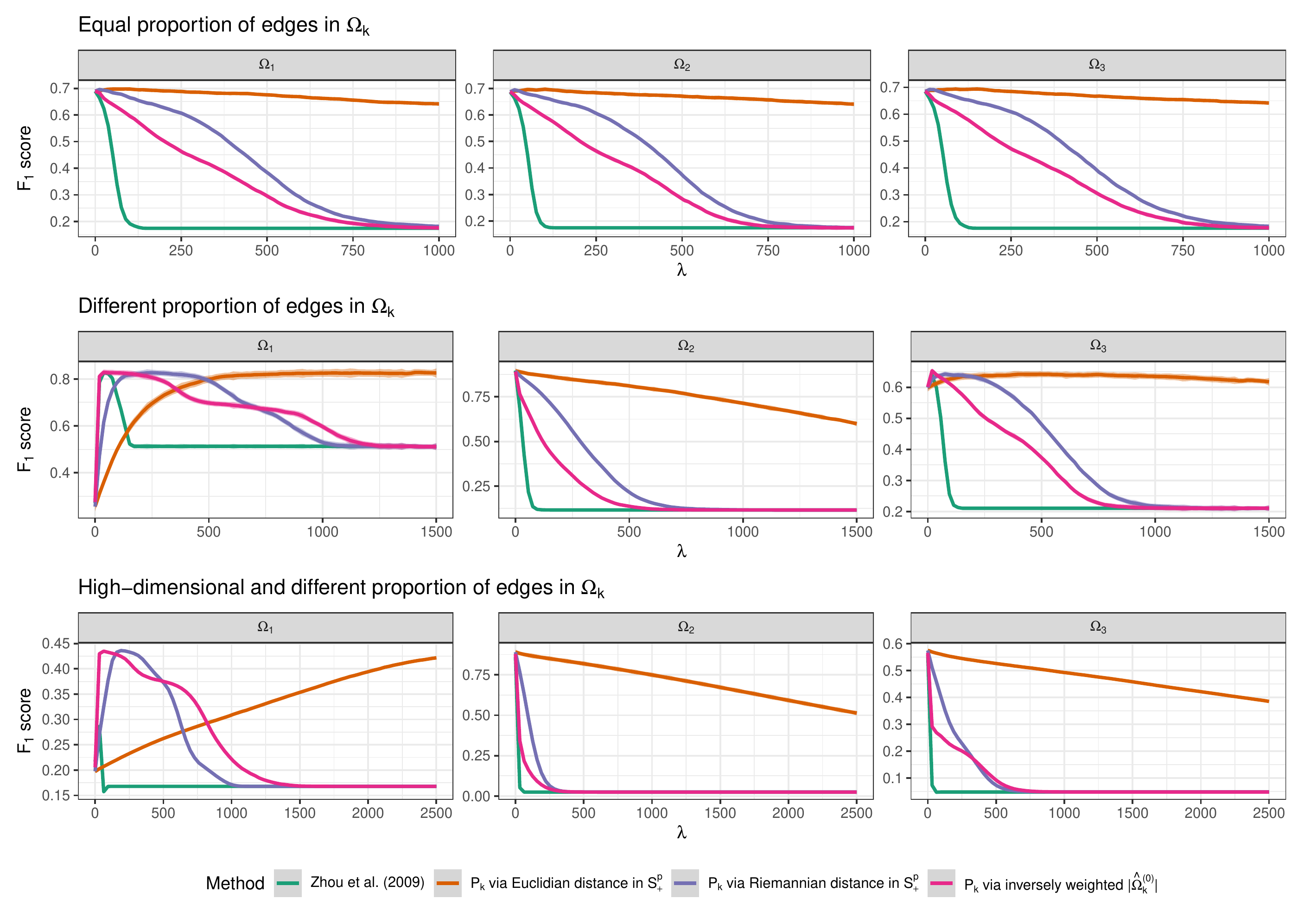}
%        \vspace*{-2cm}
    \caption{Smoothed lines plots of the $F_1$ score for $B=100$ repetitions of the simulated scenarios, varying method and shrinkage factor $\lambda$. }
  \label{fig:lambda_smooth}
\end{figure}
In all scenarios, we take equal mixing proportions $\pi_k=1/3$, $k=1,2,3$ and mean vectors equal to:
\begin{eqnarray*}
\MU_1=-1.5\mathbf{e}_{20}, \quad \MU_2=0 \mathbf{e}_{20}, \quad \MU_3=1.5\mathbf{e}_{20},
\end{eqnarray*}
for the first two scenarios, while 
\begin{eqnarray*}
\MU_1=5\mathbf{e}_{100}, \quad \MU_2=0 \mathbf{e}_{100}, \quad \MU_3=5\mathbf{e}_{100},
\end{eqnarray*}
for the high dimensional case, with $\mathbf{e}_{20}$ and $\mathbf{e}_{100}$ identifying the all-ones vector in $ \mathbb{R}^{20}$ and $ \mathbb{R}^{100}$, respectively. Such parameters induce a moderate degree of overlapping between components in the lower-dimensional case, whilst producing well-separated clusters in the high-dimensional scenario. We point out again that the primary objective of the study is assessing the recovering of the true underlying sparse precision matrices, and so we do not impose any regularization on the mean parameters. An example of the graph structures resulting from the first two scenarios are reported in Figure \ref{fig:same_p} and Figure \ref{fig:different_p} respectively. 
\begin{figure}[t]
\centering
    \includegraphics[scale=.5]{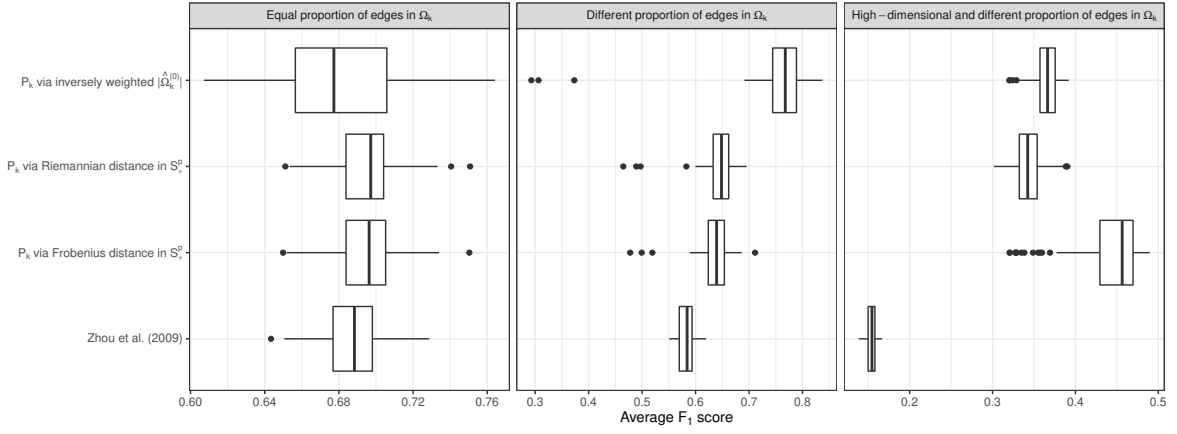}
    \caption{Boxplots of the mean $F_1$ score, averaged over $\hat{\Omega}_1$, $\hat{\Omega}_2$ and $\hat{\Omega}_3$, for the $B=100$ simulations in the three considered scenarios. For each simulation and method, the shrinkage parameter $\lambda$ is selected according to the modified BIC defined in \eqref{eq:modified_BIC}.}
  \label{fig:box_BIC}
\end{figure}
We repeat the experiment $B=100$ times, and for each replication we fit the model in Equation \eqref{eq:obj_function}, computing $\mathbf{P}_k$'s as follows:
\begin{itemize}
\item \textit{Zhou et al. (2009)}: $\mathbf{P}_k$'s set equal to the all-one matrix for  $k=1,2,3$
   \item  \textit{$\mathbf{P}_k$ via Frobenius distance in $\mathbb{S}^p_+$}: $\mathbf{P}_k$'s computed as in \eqref{weigthed_by_distance_to_diagW0}, with $\mathcal{D}(\cdot,\cdot)$ the Frobenius distance in the $\mathbb{S}^p_+$ space, %\comment{MF- I think the correct definition of Euclidean distance for matrices is Frobenius distance, we should change it in the figures}
   \item     \textit{$\mathbf{P}_k$ via Riemannian distance in $\mathbb{S}^p_+$}: $\mathbf{P}_k$'s computed as in \eqref{weigthed_by_distance_to_diagW0}, with $\mathcal{D}(\cdot,\cdot)$ the Riemannian distance in the $\mathbb{S}^p_+$ space,
   \item \textit{$\mathbf{P}_k$ via inversely weighted $|\hat{\OMEGA}^{(0)}_k|$}: $\mathbf{P}_k$'s computed as in \eqref{weigthed_by_w0}.
\end{itemize}
A grid of equispaced $100$ elements for the penalty term $\lambda$ is considered, with lower and upper extremes set to:
\begin{eqnarray} \label{eq:lambda:range}
\left[ 0,  \max_k \left\{ \max \left\{ |\mathbf{S}^{(0)}_k-\mathbf{I}_p|\right\}\dfrac{n_k^{(0)}}{2}\right\} \right],
\end{eqnarray}
where the inner max operation is taken element-wise, and with $\mathbf{S}_k^{(0)}$ and $n_k^{(0)}$ the starting estimates of the sample covariance matrices and their associated sample sizes, initialized via Gaussian mixture models provided by the \texttt{mclust} software \citep{Scrucca2016}.
Other initialization strategies are clearly possible, as described in Section \ref{sec:init_omega_0}. We note that the upper term in \eqref{eq:lambda:range} forces the final estimates  $\hat{\OMEGA}_k$ to be approximately diagonal when $\mathbf{P}_k$'s are equal to all-ones matrices \citep{Zhao2012}. 

The performance of the methods, in relation to the different specification of the $\mathbf{P}_k$'s matrices, is evaluated via component-wise $F_1$ scores (see Equation \ref{eq:f1_measure}), where the problem of matching the estimated clustering to the actual classification is dealt with by means of the \texttt{matchClasses} routine of the \texttt{e1071 R} package \citep{e1071}. Simulation results are reported in the next subsection.

\subsection{Simulation results: recovering the association structure} \label{sec:sim_ass_structure}
Results for the simulated experiments are summarized in Figure \ref{fig:lambda_smooth}, depicting smoothed lines plots of the $F_1$ scores for the estimated $\OMEGA_k$, $k=1,2,3$, under the three considered scenarios, for different specifications of the matrices $\mathbf{P}_k$'s and shrinkage factor $\lambda$. By visually exploring Figure \ref{fig:lambda_smooth}, several interesting patterns emerge.

First off, it is immediately noticed that the methods performance in the first scenario does not vary across components, with strong similarities between $F_1$ score trajectories for $\OMEGA_1$, $\OMEGA_2$ and $\OMEGA_3$. This is expected, as each precision matrix is generated with a probability of connection equal to $0.5$. This is the ``gold-standard'' scenario for the method described in \cite{zhou2009penalized}, since in principle a common $\lambda$ should be sufficient for achieving the same group-wise degree of sparsity. Indeed, the highest $F_1$ values (around $0.7$) are achieved by all proposals when small penalty values are considered. Notwithstanding, multiplying the common shrinkage term by a group-specific factor $\mathbf{P}_k$ moderates the rapid decline in performance when $\lambda$ increases. Particularly, computing the $\mathbf{P}_k$'s as a function of the Frobenius distance between $\hat{\OMEGA}^{(0)}_k$ and ${\rm diag}\left(\hat{\OMEGA}^{(0)}_{k}\right)$ greatly downweights the impact of the common penalty term, making the procedure less sensitive to the selection of  $\lambda$.

The beneficial effect of group-wise $\mathbf{P}_k$'s becomes apparent in the second scenario, where the true precision matrices have dissimilar a priori probability of connection. $F_1$ trajectories are component-wise different: the almost diagonal $\OMEGA_1$ would require a higher shrinkage for recovering the very sparse underlying graph structure, whereas the highly connected $\OMEGA_2$ is well-estimated when low values of $\lambda$ are considered. This trade-off is mitigated by the $\mathbf{P}_k$ factor, which adjusts for under or over-connectivity within the estimation process. In particular, every suggested approach succeeds in improving the strategy of \cite{zhou2009penalized}, with $F_1$ patterns for our proposals almost always dominating the common shrinkage method. This behavior is intensified even further in the scenario with a larger number of variables, where as soon as $\lambda$ increases, the proportion of incorrectly missed edges produces a huge drop in the $F_1$ score for the second component. As previously highlighted, the \textit{$\mathbf{P}_k$ via Frobenius distance in $\mathbb{S}_+^p$} is the solution enforcing the greatest discount on the common $\lambda$, greatly improving the results for $\OMEGA_2$ at the expense of overestimating the true number of edges for $\OMEGA_1$.

Figure \ref{fig:lambda_smooth} shows the overall superiority of our proposals with respect to a common penalty framework in group-wise recovering of sparse precision matrices. Nevertheless, when it comes to performing the analysis, a single value of $\lambda$ must be chosen. We make use of the BIC criterion defined in Section \ref{sec:further_aspect} to determine the best $\lambda$ for each method and instance of the simulated experiments. For the $B=100$ simulations in the three scenarios, the resulting empirical distribution of the mean $F_1$ score averaged over $\hat{\OMEGA}_1$, $\hat{\OMEGA}_2$, and $\hat{\OMEGA}_3$ is reported in Figure \ref{fig:box_BIC}. %For each simulation and method, $\lambda$ providing the highest value of \eqref{eq:modified_BIC} has been selected.
As expected, the overall results do not dramatically change in the \textit{equal proportion of edges in $\OMEGA_k$} case. On the other hand, for the other two scenarios, our proposals perform substantially better compared to \citet{zhou2009penalized}, especially in the case with larger number of variables.  None of the introduced methods for computing $\mathbf{P}_k$'s seems to outperform the others; nonetheless, it is clear that, whenever the degree of conditional dependence varies greatly across components, weighting the common penalty $\lambda$ by a group-specific factor improves the recovering of the group-wise different sparse structures.

\subsection{Simulation results: clustering performance} \label{sec:ari_test}
\begin{figure}[t]
\centering
    \includegraphics[scale=.6]{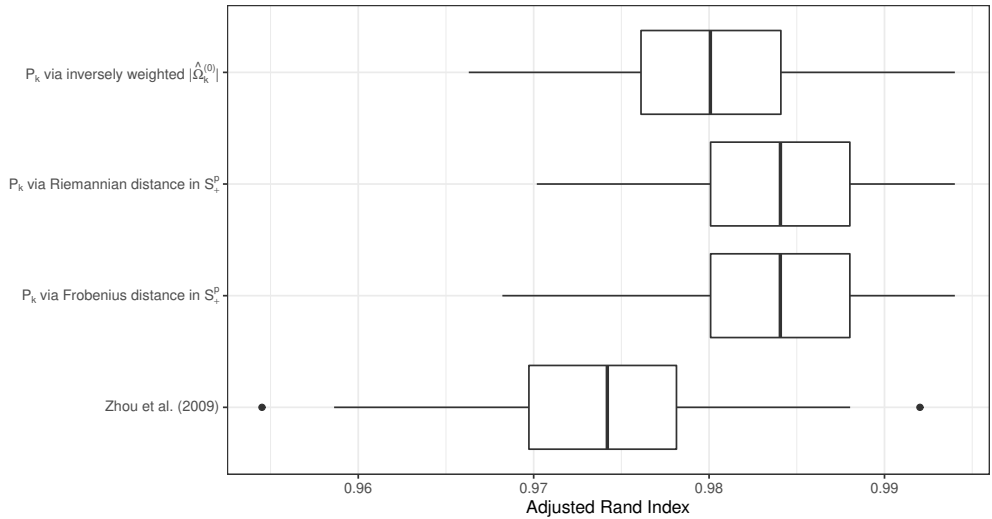}
    \caption{Boxplots of the Adjusted Rand Index for the $B=100$ simulations of the scenario described in Section \ref{sec:ari_test}. For each method, the shrinkage parameter $\lambda$  and the number of components $K$ are selected according to the modified BIC defined in \eqref{eq:modified_BIC}.}
  \label{fig:box_ARI}
\end{figure}

The previous section showcases the ability of our proposals in learning  group-wise different sparse structures in the components precision matrices. In doing so, we did not directly assess the obtained clustering, as the mean vectors induced adequately well-separated components. We hereafter evaluate the recovering of the underlying data partition by 
generating further $B=100$ samples from the \textit{Different proportion of edges in $\OMEGA_k$} scenario. Contrarily to the previous study, we fix $\MU_1=\MU_2=\MU_3=\mathbf{0}$. That is, components differ only on the basis of their conditional dependence structures, and thus the final allocation is entirely driven by the estimated precision matrices $\hat{\OMEGA}_k$. In this case, where both the clustering and the association structures are of interest, a comprehensive model selection strategy is needed to choose $K$ and to properly tune $\lambda$. More specifically, coherently with both the model-based clustering literature and the penalized estimation schemes setting, we evaluate each model on a grid of $\lambda$ values, whose range is computed as in \eqref{eq:lambda:range}, and for different mixture components $K\in\{2,3,4,5\}$. For each model, we select $\lambda$ and $K$ according to the BIC criterion defined in \eqref{eq:modified_BIC}. The Adjusted Rand Index \citep[ARI,][]{hubert:arabie:1985} is employed for comparing the estimated classification with the true data partition. The results are reported in Figure  \ref{fig:box_ARI}. While the resulting clustering is satisfactorily close to the true one for all models, including in the penalty specification the group-wise shrinkage matrices $\mathbf{P}_k$'s seems to improve the overall performance. Careful analysis of the results demonstrate that the model with common penalty struggles in separating the components with high and medium degree of connectivity. Such a behavior is exacerbated even further when the data dimensionality is equal or even larger than the sample size, as demonstrated in the next section.

\subsection{Simulation results: $p \geq n$ scenarios} \label{sec:p_bigger_n}
\begin{figure}[t]
\centering
    \includegraphics[scale=.7]{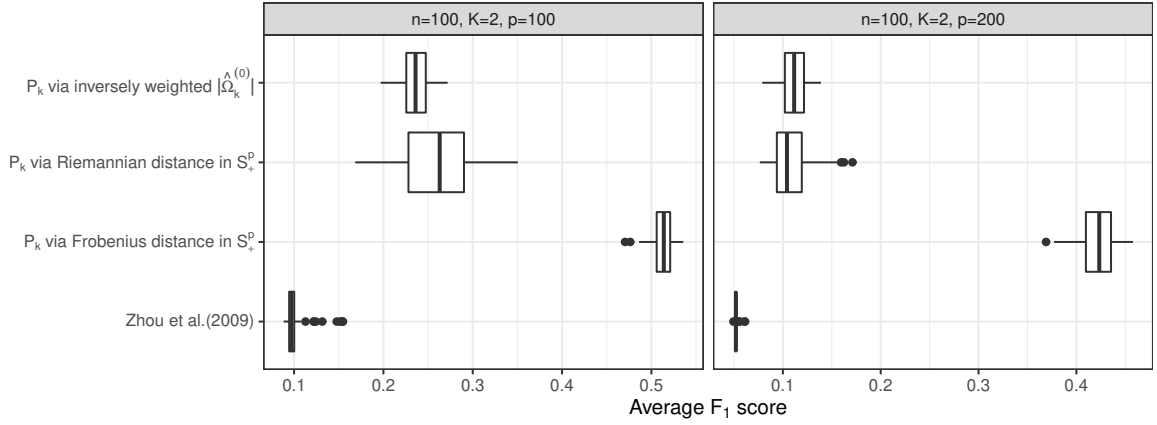}
    \caption{Boxplots of the mean $F_1$ score, averaged over $\hat{\Omega}_1$, $\hat{\Omega}_2$, for the $B=100$ simulations in the two $p \geq n$ scenarios. For each simulation and method, the shrinkage parameter $\lambda$ is selected according to the modified BIC defined in \eqref{eq:modified_BIC}.}
  \label{fig:box_F1_p_bigger_n}
\end{figure}

\begin{figure}[t]
\centering
    \includegraphics[scale=.7]{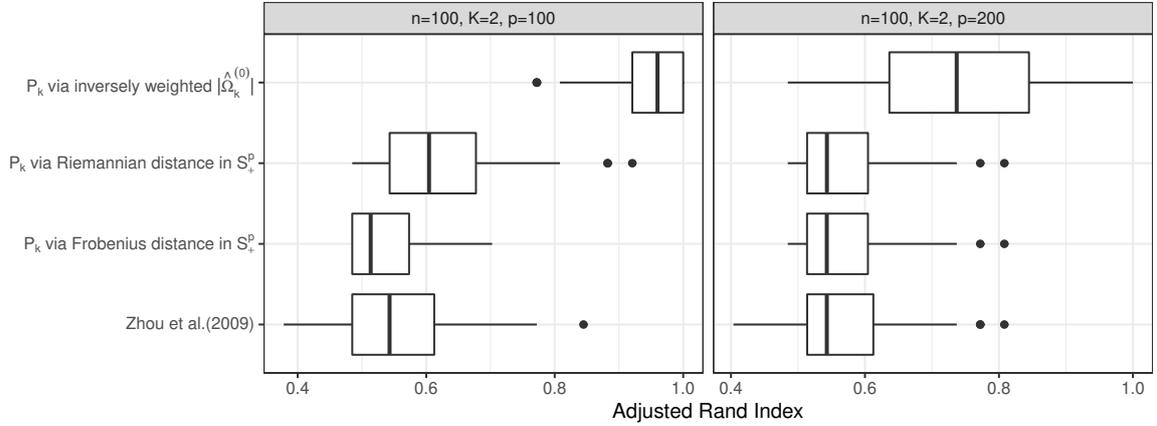}
    \caption{Boxplots of of the Adjusted Rand Index for the $B=100$ simulations in the two $p \geq n$ scenarios. For each simulation and method, the shrinkage parameter $\lambda$ is selected according to the modified BIC defined in \eqref{eq:modified_BIC}.}
  \label{fig:box_ARI_p_bigger_n}
\end{figure}

%In the two previous subsections we have thoroughly assessed the performance of our proposals in recovering the group-wise association structures and the true underlying data partition. We now
In this section we further explore, via simulation, the applicability of our proposals when the data dimension is equal or even larger than the sample size. In details, the following data generating process is considered: we sample $n=100$ observations from a $K=2$ Gaussian mixture, with precision matrices generated according to a sparse at random Erd\H{o}s-R\'{e}nyi graph structure with probabilities of connection equal to $0.1$ and $0.8$ (the same structure as for the first two components in the \textit{Different proportion of edges in $\OMEGA_k$} scenario, see Section \ref{sec:exp_setup}). We contemplate two different cases, setting the number of variables equal to $p=100$ and $p=200$. For each scenario, we replicate the experiments $B=100$ times, monitoring the resulting mean $F_1$ scores and Adjusted Rand Index: Figures \ref{fig:box_F1_p_bigger_n} and \ref{fig:box_ARI_p_bigger_n} report the resulting boxplots for the former and latter metric, respectively.

Similarly to what observed in the right-most panel of Figure \ref{fig:box_BIC}, the inclusion of data-driven $\mathbf{P}_k$'s assures a better recovery of the true conditional association structure: the  mean $F_1$ scores displayed by our proposals are significantly higher than the one displayed by \citet{zhou2009penalized}. Particularly, employing the strategy described in Section \ref{sec:weigthed_by_distance_to_diagW0}, coupled with a Frobenius distance, seems to outperform all the other procedures. Interestingly, our proposals showcase fairly good results even in the challenging $n=100,\:K=2,\:p=200$ scenario, demonstrating the effective applicability of such criteria even when the feature space is bigger than the sample size. The same holds only partially true when we monitor the Adjusted Rand Index (Figure \ref{fig:box_ARI_p_bigger_n}). As it may be expected, the high dimensionality deteriorates the recovery of the true data partition for all penalized models. Nonetheless, the ARI for methods with group-wise different $\mathbf{P}_k$'s is never lower than the one of \citet{zhou2009penalized}. Moreover, the  clustering retrieved by the \textit{$\mathbf{P}_k$ via inversely weighted $|\hat{\OMEGA}^{(0)}_k|$} procedure is significantly better than all the other alternatives, particularly for the $n=100,\:K=2,\:p=100$ scenario.

\subsection{A note on computing times} \label{sec:comp_times}
All the penalized methods considered in the aforementioned simulation studies rely on an iterative algorithm when performing parameters estimation. To this extent, it is of interest to investigate the required elapsed time to fit the models. Table \ref{tab:comp_analysis} reports the average computing times and associated standard deviations for different methods and simulated scenarios. All the simulated experiments were run on a computer cluster with 32 processors Intel Xeon E5-4610 @2.3GHz.
\begin{table}[t]
\centering
\caption{Average computing time (in seconds) over $B = 100$ runs for the simulation studies reported in Sections \ref{sec:sim_ass_structure} (first three columns) and  \ref{sec:p_bigger_n} (last two columns). The metric refers to the average time required in fitting a model with a single shrinkage term $\lambda$. Standard errors are reported in parentheses.}
\label{tab:comp_analysis}
\hspace*{-1cm}\begin{tabular}{llllll}
  \hline
% & Equal proportion of edges in $\OMEGA_k$ & Different proportion of edges in $\OMEGA_k$ & High dimensional and different proportion of edges in $\OMEGA_k$ & $n=100, K=2, p=100$ & $n=100, K=2, p=200$ \\ 
  & Equal prop of & Diff prop of & High dim and diff & Section \ref{sec:p_bigger_n} & Section \ref{sec:p_bigger_n}  \\ 
    & edges in $\OMEGA_k$ & of edges in $\OMEGA_k$ &  prop of edges in $\OMEGA_k$ & $p=100$  & $p=200$ \\ 
  \hline
Zhou et al.(2009) & $0.0918$ & $0.0892$ & $0.1461$ & $1.5895$ & $10.5871$ \\ 
 & $(0.073)$ & $(0.106)$ & $(0.055)$ & $(0.092)$ & $(0.495)$ \\ 
 \\
  $P_k$ via Frobenius  & $0.0352$ & $0.0458$ & $2.7869$ & $3.024$ & $22.2367$ \\ 
distance in $S^p_+$ & $(0.016)$ & $(0.098)$ & $(0.723)$ & $(2.514)$ & $(18.184)$ \\ 
 \\
  $P_k$ via Riemannian  & $0.0325$ & $0.0378$ & $0.4767$ & $1.7297$ & $11.7203$ \\ 
distance in $S^p_+$ & $(0.014)$ & $(0.05)$ & $(0.532)$ & $(0.559)$ & $(4.547)$ \\ 
 \\
  $P_k$ via inversely  & $0.0294$ & $0.0347$ & $0.2065$ & $1.5943$ & $10.7254$ \\ 
weighted $|\hat{\Omega}^{(0)}_k|$ & $(0.015)$ & $(0.045)$ & $(0.156)$ & $(0.118)$ & $(0.915)$ \\ 
   \hline
\end{tabular}
\end{table}

First off, recall that the calculation of the group-wise different $\mathbf{P}_k$'s in our proposals is performed only once prior to start the EM algorithm. Therefore, at least in principle, the extra computational effort required by our methods with respect to the one by \citet{zhou2009penalized} amounts only to compute the $\mathbf{P}_k$'s at the beginning of the iterative procedure. By looking at Table \ref{tab:comp_analysis} we notice that, irrespective of the methods and quite naturally, an higher dimensionality is associated with a longer computational time. At any rate, in low dimensional settings (\textit{Equal proportion of edges in $\OMEGA_k$} and \textit{Different proportion of edges in $\OMEGA_k$}) our methods seems to be even (slightly) faster than \citet{zhou2009penalized}. Conversely, when the dimensionality increases,  \citet{zhou2009penalized} showcases, as expected, the shortest computing times. This is particularly true if compared with the \textit{$P_k$ via Frobenius distance in $S^p_+$ } strategy, for which it appears a larger number of EM iterations are necessary to reach convergence in high-dimensional settings. Notwithstanding we argue that, as extensively demonstrated  in the previous sections, the small extra price to pay in terms of computing time tends to be well worth when it comes to detect clusters with diverse degrees of sparsity between the variables.

All in all, considering a group-wise penalty not only improves the estimation of the component precision matrices, but it also enhances the quality of the resulting data partition in all the simulated scenarios, even in the most challenging ones in which the data dimension is equal or even greater than the sample size. The same happens in the illustrative data examples, as it is reported in the next section. 

\color{black}
\section{Illustrative datasets} \label{sec:true_data}
This section presents two illustrative data examples. %We perform model based clustering with sparse precision matrices and compare different methods in defining the penalty matrices $\mathbf{P}_k$s.
In this case, contrarily to the synthetic experiments reported in Section~\ref{sec:application}, the true underlying graph structure is unknown %and cannot be recovered
and thus we employ several metrics to assess the model performance under the different definitions of the $\mathbf{P}_k$'s matrices. Classification accuracy is evaluated as usual via the ARI, while the number of non-zero estimated parameters in the precision matrices (indicated by $d_{\OMEGA}$) is taken as a proxy of model complexity. The identification of the underlying conditional association structure is evaluated by implementing the following approach. Using the true class labels %(which are known in the following benchmark datasets),
we compute the data class-specific precision matrices; then, for each method, we measure the median %and median
distance (in terms of the Frobenius norm) between the empirical and the estimated sparse precision matrices. In doing so, we make again use of the  \texttt{matchClasses} routine to match the empirical class-specific precision matrix with its corresponding sparse estimate. In detail, the \textit{Median Frobenius Distance} metric (MFD) is computed as follows:
\begin{equation} \label{eq:av_frob_dist}
\underset{k \in 1,\ldots, K}{\operatorname{median}}\left(||\hat{\OMEGA}_k - \bar{\OMEGA}_k||_F\right)
\end{equation}
with $||\cdot||_F$ denoting the Frobenius norm, while $\hat{\OMEGA}_k$ is the estimated precision matrix of the $k$-th component for a given method and $\bar{\OMEGA}_k$ is the empirical $k$-th class precision matrix, computed using the true labels.

%precision matrix with the corresponding estimated sparse precision matrix
% between Frobenius norm across all clusters
%
% we match the estimated clustering with the actual classification of the observations, and consequently also matching the empirical class-specific precision matrix with the corresponding estimated sparse precision matrix. We then compute 
%
%To assess the quality of the results we . We evaluate the 
Similarly to the first simulation study, we fix the number of clusters to the number of classes available in the data; we do so in order to focus the attention on the model selection aspect concerning the recovery of the conditional association structure, rather than on the selection of the number of components in the mixture. Lastly note that data have been standardized before applying any modeling procedure, as it is customarily done with penalized estimation. Nonetheless, we acknowledge that standardization can have an impact on the results and we refer to the recent work by \citet{carter2021partial} for a thorough discussion.

%we evaluate the ability by adopting the following procedure. Using the class labels (which are known in the following benchmark datasets), we compute the data class-specific precision matrices. Then, for each method, we match the estimated clustering with the actual classification of the observations, and consequently also matching the empirical class-specific precision matrix with the corresponding estimated sparse precision matrix. We then compute the average and median Frobenius norm across all clusters, as a measure of . Also BIC value and number of covariance parameters are reported.

\subsection{Olive Oil} \label{sec:olive_oil}
The first dataset reports the percentage composition of $p=8$ fatty acids in $n=572$ units of olive oil. The oil samples come from $K=9$ Italian regions: the aim is to recover the geographical partition of the oils by means of their lipidic features. This dataset was firstly described in \citet{forina1983classification} and it is available in the R package \texttt{pgmm} \citep{pgmm2018}. Results for the considered methods are reported in Table \ref{tab:olive_analysis}. 

\begin{figure}[t]
\centering
    \includegraphics[scale=1]{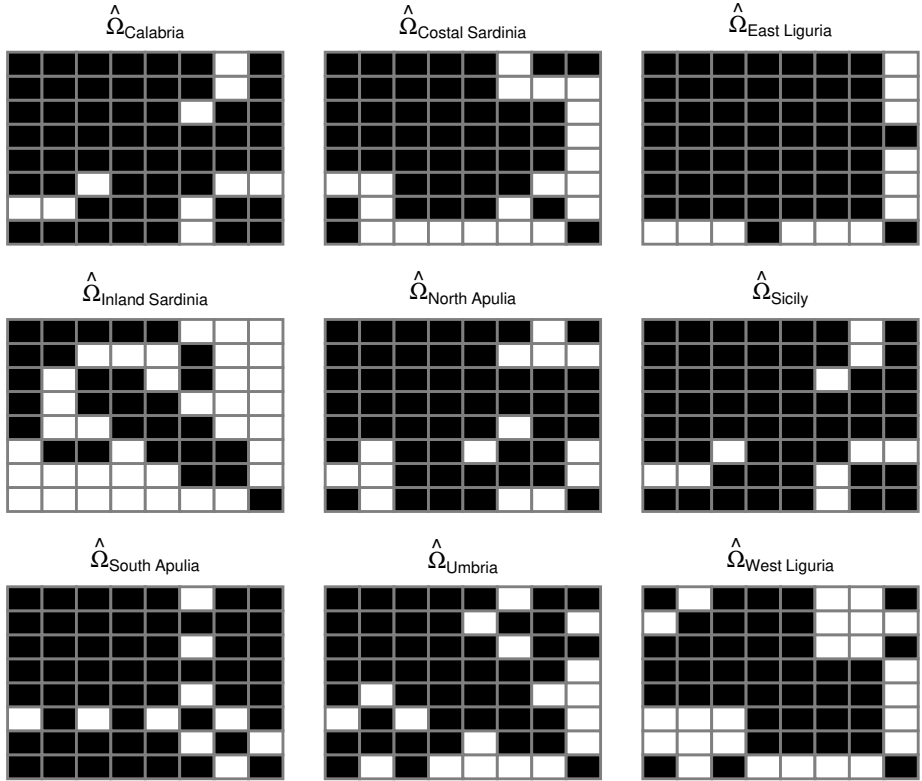}
    \caption{Estimated graphs in the precision matrices for the \textit{$\mathbf{P}_k$ via inversely weighted $|\hat{\OMEGA}^{(0)}_k|$} approach. Black squares denote the presence of an edge between the two variables. Olive oil dataset.}
  \label{fig:olie_oil}
\end{figure}

Together with the different specification of $\mathbf{P}_k$'s for sparse estimation, we include in the comparison the standard model-based clustering approach with eigen-decomposed covariance matrices selected using BIC, fitted via the \texttt{mclust} software \citep{Scrucca2016}.
% the benchmark parsimonious (unpenalized) model-based clustering approach, fitted via the \texttt{mclust} software \citep{Scrucca2016}. 
\begin{table}[bt]
\centering
\caption{BIC, Adjusted Rand Index, number of estimated parameters and Median Frobenius Distance, as defined in \eqref{eq:av_frob_dist}, for different model-based clustering methods. Olive oil dataset.}
\label{tab:olive_analysis}
\begin{tabular}{rrrrr}
  \hline
 & BIC & ARI & $d_{\OMEGA}$ & MFD \\ 
  \hline
\texttt{mclust} VVE & -4790 & 0.6586 & 100 & 758 \\ 
  Zhou et al.(2009) & -5302 & 0.6724 & 320 & 830 \\ 
  $\mathbf{P}_k$ via inversely weighted $|\hat{\Omega}^{(0)}_k|$ & -5058 & 0.7199 & 242 & 421 \\ 
  $\mathbf{P}_k$ via Frobenius distance in $\mathbb{S}^p_+$ & -5286 & 0.6875 & 312 & 701 \\ 
  $\mathbf{P}_k$ via Riemannian distance in $\mathbb{S}^p_+$ & -5282 & 0.6812 & 314 & 798 \\ 
   \hline
\end{tabular}
\end{table}
For all penalized methods, the selection criterion defined in \eqref{eq:modified_BIC} is used to identify the best $\lambda$ in a data-driven fashion. In general, penalized models outperform \texttt{mclust} VVE (different volume and shape but same clusters orientation) in recovering the true data partition. 
This might be due to the rigid dependence structure imposed by such model, where the association among variables is forced to be equal across all components.
%This is due to the rigid dependence structure imposed by such model, forcing it to be equal across all components. 
Notice that including a data-dependent specification for $\mathbf{P}_k$'s slightly improves the clustering accuracy with respect to the all-one matrix \citep{zhou2009penalized}. Moreover, the overall model complexity is reduced: the method with common penalty selects a $\lambda$ that induces a mild sparsity, as a total of $Kp(p+1)/2=324$ parameters would be considered in a fully-unconstrained estimation. On the other hand, for our proposals the number of non-zero inverse covariance parameters $d_{\OMEGA}$ is lower than for the full model, and particularly the \textit{$\mathbf{P}_k$ via inversely weighted $|\hat{\OMEGA}^{(0)}_k|$} approach substantially reduces the number of estimated parameters, whilst showcasing the highest ARI and the lowest Median Frobenius distance. The corresponding graphs for the $9$ different clusters are reported in Figure \ref{fig:olie_oil}, in which we see that the conditional dependence structure appreciably varies  across regions, with our proposal taking advantage of it in the estimation phase.

\subsection{Handwritten digits recognition} \label{sec:digit_recogn}
\begin{figure}[t]
\centering
    \includegraphics[scale=.6]{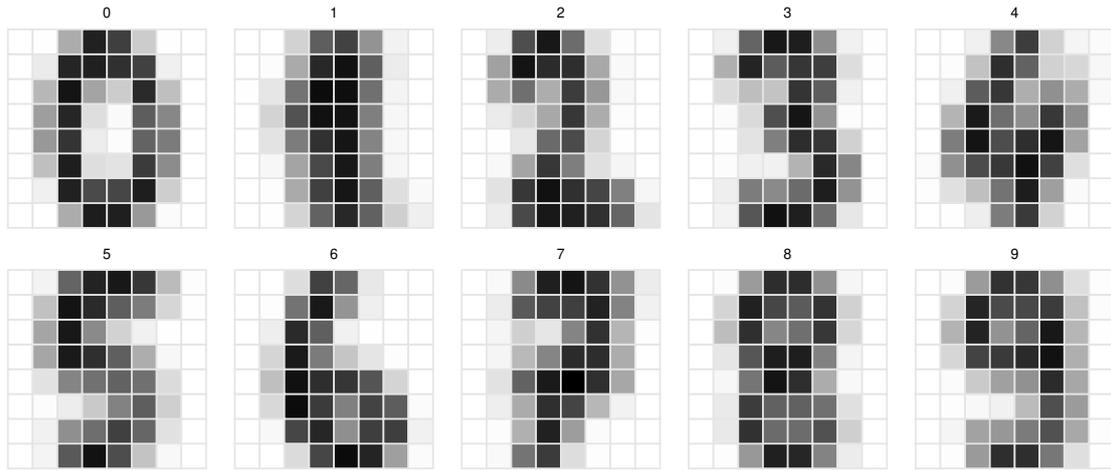}
    \caption{Image representation of the means of digits $0$ to $9$. Handwritten digits dataset.}
  \label{fig:avg_digits}
\end{figure}
\begin{figure}
\centering
    \includegraphics[scale=.8, angle=90]{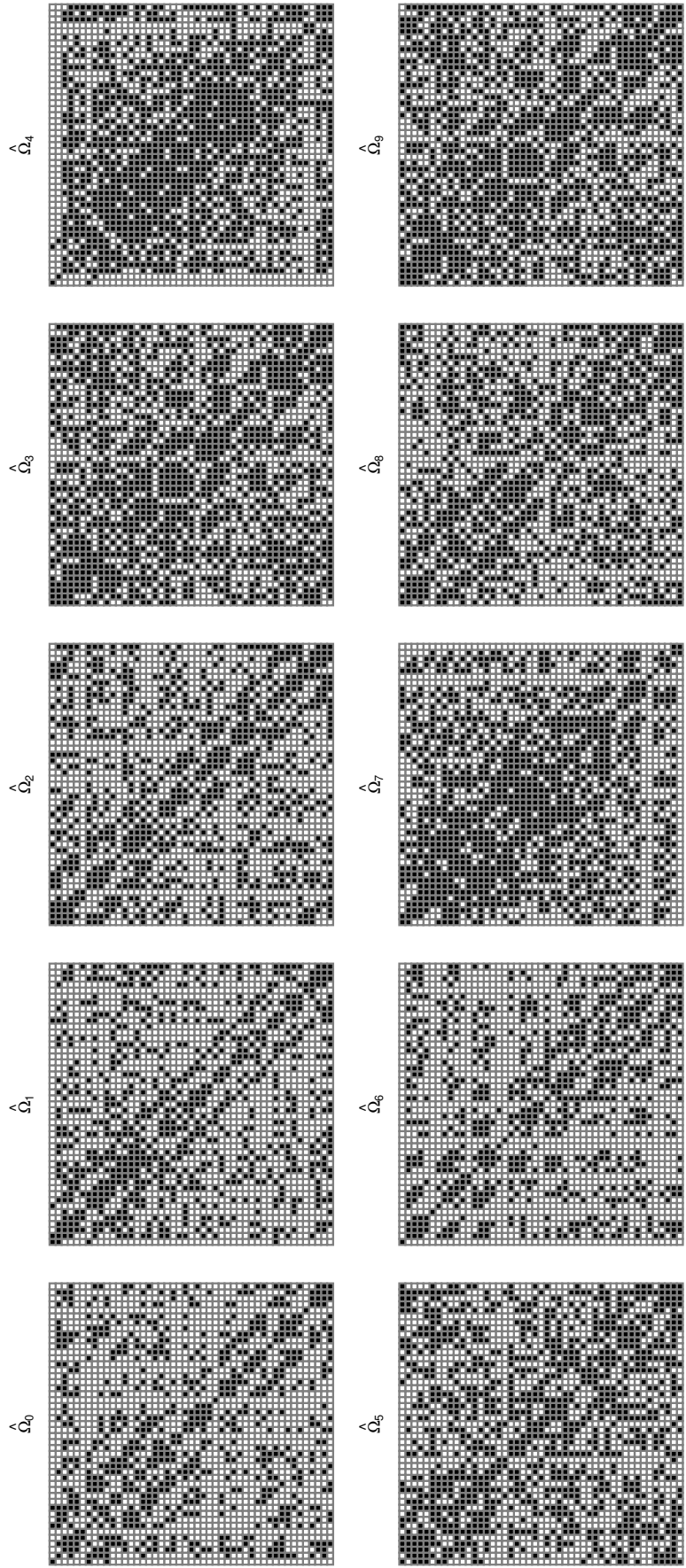}
    \caption{Estimated graphs in the precision matrices for the \textit{$\mathbf{P}_k$ via Riemannian distance in $S^p_+$ } approach. Black squares denote the presence of an edge between the two variables. Handwritten digits dataset.}
  \label{fig:omega_digits}
\end{figure}

The second dataset, publicly available in the University of California Irvine Machine Learning data repository (\texttt{http://archive.ics.uci.edu/ml/datasets/optical+\linebreak recognition+of+handwritten+digits}), contains $n=5620$ samples of handwritten digits represented by $64$ features. Each variable counts the pixels of a $ 8\times8$ grid in which the original images were divided. 
%The original $32 \times 32$ images were divided into $4 \times 4$ nonoverlapping blocks
%Each variable counts the on pixels of the $4 \times 4$ nonoverlapping blocks in which the original $32 \times 32$ images were divided.
The aim is to recognize the $K=10$ digits by means of the penalized procedures introduced in the paper. This clustering problem is more challenging than the one presented in Section \ref{sec:olive_oil}, due to both the higher dimensionality and the narrower separation between classes (see Figure \ref{fig:avg_digits}). Before applying the different clustering methods, we employ a preprocessing step, excluding from the subsequent analysis all predictors with near zero variance. This boils down to essentially remove the left-most and right-most pixels in each image, as being mostly white they contain no separating information. To this task, we use the default routines available in the R package \texttt{caret} \citep{caret}.  After having eliminated these variables, we are left with $p=47$ features, which are then considered to perform model-based clustering. Results are reported in Table \ref{tab:digits_analysis}.
\begin{table}
\centering
\caption{BIC, Adjusted Rand Index, number of estimated parameters and Median Frobenius Distance, as defined in \eqref{eq:av_frob_dist}, for different model-based clustering methods. Handwritten digits dataset.}
\label{tab:digits_analysis}
\begin{tabular}{rrrrr}
  \hline
 & BIC & ARI & $d_{\OMEGA}$ & MFD \\ 
  \hline
\texttt{mclust} EEE & -521220 & 0.6489 & 1128 & 172 \\ 
  Zhou et al.(2009) & -388862 & 0.6837 & 4914 & 148 \\ 
  $\mathbf{P}_k$ via inversely weighted $|\hat{\OMEGA}^{(0)}_k|$ & -368604 & 0.6820 & 3436 & 104 \\ 
  $\mathbf{P}_k$ via Frobenius distance in $\mathbb{S}^p_+$ & -391359 & 0.6827 & 6066 & 146 \\ 
  $\mathbf{P}_k$ via Riemannian distance in $\mathbb{S}^p_+$ & -388902 & 0.6841 & 5206 & 147 \\ 
   \hline
\end{tabular}
\end{table}
The parsimonious structure selected by \texttt{mclust} forces the precision matrices to be all equal across groups. This rigid constraint undermines the classification accuracy and the uncovering of the conditional dependence structure, resulting in the worst ARI and Median Frobenius distance metrics. Conversely, the penalized methods are able to shrink the estimates in a group-wise manner. This is especially true in our proposals for which, even though the resulting classification accuracy is not dramatically affected, the Median Frobenius distance is always smaller than \citet{zhou2009penalized}. In Figure \eqref{fig:omega_digits} we report the estimated graphs in the precision matrices for the \textit{$\mathbf{P}_k$ via Riemannian distance in $S^p_+$ } approach which results in the highest ARI.
%\textcolor{red}{MF -- I think we should report the plots for the method inversely weighted, as it has the largest BIC. AC: eviterei di far vedere inversely weighted perchè non si vedono differenze in termini di numero di edges tra le varie cifre, quindi un plot del genere rema contro la finalità dell'articolo :-(. }
Lastly note how the number of estimated edges appreciably differ between digits, an aspect that is implicitly taken into account in our data-driven specification of the $\mathbf{P}_k$'s matrices.

\section{Discussion}\label{sec:discussion}

The present paper has highlighted the limitations of imposing a single penalty when performing sparse estimation of component precision matrices in a multiclass setting. We have argued that methods enforcing similarities in the graphical models across groups may not be adequate for classification, since they have detrimental effects when it comes to groups discrimination, in particular in the case of clustering. Thus, we have focused our attention on the penalized model-based method with sparse precision matrices framework of \citet{zhou2009penalized}, where class-specific differences are preserved. Nonetheless, this methodology does not account for situations in which a component displays under or over-connectivity with respect to the remaining ones. To this extent, we have proposed some procedures to incorporate group-specific differences in the estimation, enforcing a carefully initialized solution to drive the algorithm in under or over penalizing specific components. Numerical illustrations and analyses on real data have confirmed the validity of our proposals. By means of our solutions we have achieved both group-wise flexibility in the precision matrices reconstruction and we have mitigated the impact the common shrinkage factor has in the overall sparse estimation.

% AC: I think this is redundant here. Note that the primary aim of this work has been to bring to light a shortcoming of the available multiclass precision matrix estimation approach and to propose some viable solutions to overcome it. Nonetheless, the proposals we have discussed are not meant to be a panacea useful for all the possible settings as other plausible strategies might be fruitfully adopted. In this context, a work that is worth mentioning is the one by \citet{fop:2019}, where the authors consider a purpose-dependent and group specific penalization terms in order to parsimoniously characterize the marginal association among the variables in a Gaussian mixture modelling setting. 
The present paper opens up a quite natural direction for future research: the penalized approach could be adapted to estimate sparse covariance matrices, rather than precision matrices. In the Gaussian case, a missing edge between two nodes in the \emph{Gaussian covariance graph model} corresponds to two variables being marginally independent, and the so called \emph{covariance graph} \citep{chaudhuri:2007} allows to represent the pattern of zeroes in the covariance matrices. A related methodology based on cluster-specific penalties has been recently introduced by \cite{fop:2019}, unfortunately, such an approach relies on a time-consuming graph structure search, making it less attractive in high dimensional problems. On the other hand, the definition of a penalized likelihood that incorporates a covariance graphical lasso term \citep{bien:2011, Wang2014} can be effectively employed in these scenarios: model definitions are being explored and they will be the object of future work.

The framework proposed here has also interesting connections with the notion of global-local shrinkage developed in the Bayesian literature for sparsity inducing priors \citep{Bhattacharya2015,Polson2010}.  The general formulation of these priors is based on a normal scale mixture representation, where the mean is zero and the variance is expressed as the product of two nonnegative parameters: one scaling parameter pulls the global shrinkage towards zero, while the other allows for modifications in the amount of shrinkage \citep{Bhattacharya2015}. Global-local shrinkage priors for Gaussian graphical models have been employed in \cite{leday2017} for gene network inference in the case of a homogeneous population. The authors develop a simultaneous equations modeling approach for graph inference, where the regression parameters are given Gaussian scale mixture priors for local and global shrinkage, which allows borrowing of information among the regressions and encourages the posterior expectation of the corresponding entries of the precision matrix to be shrunk towards zero. As pointed out by \cite{leday2017}, compared to \cite{meinshausen:2006}, a disadvantage of these priors is that they do not automatically perform variable selection, hence the graph structure needs to be recovered by thresholding of the posterior means of the regression coefficients. An alternative use of global-local shrinkage priors is in the Bayesian graphical lasso of \cite{wang2012}. Here, it is shown that the graphical lasso estimator is the maximum a posteriori of a Bayesian hierarchical model where the entries of the precision matrix have exponential and double exponential prior distributions, which can be represented as a scale mixture of normals. We note that the graphical modeling Bayesian frameworks of \cite{leday2017} and \cite{wang2012} are developed for the case of a homogeneous sample, and that global-local shrinkage is intended only in terms of joint shrinkage of all the entries of $\OMEGA$, allowing only for variable and scale-specific adjustments. In contrast, in our proposed approach, global-local shrinkage would be intended in terms of joint shrinkage of the component precision matrices $\OMEGA_k$ towards a common level of sparsity, with cluster related adaptations. The penalty term $\lambda$ could be considered the global shrinkage factor, which equally shrinks the entries of the precision matrices across the mixture components, while the weighting matrices $\mathbf{P}_k$'s allow for local cluster-specific adjustments. Following the literature on penalized model-based clustering, we devised our proposal under a penalized likelihood framework, which has computational advantages especially in high-dimensional scenarios. However, with the purpose of sparse Bayesian model-based clustering, carefully defined prior distributions could be defined for global-local shrinkage across mixture components and within clusters; these considerations open a path for future developments of our proposal in a Bayesian context and are currently under exploration.

As a last worthy note, even if the proposed procedure is applicable in a general setting, we believe that the definition of group-specific penalties should never leave aside prior information and subject-matter knowledge whenever available, as their incorporation in the methodology can be strongly beneficial for the analysis.

%--------------------------------------------------------------------------
\bibliographystyle{apalike}
\bibliography{biblio.bib}

\end{document}